%% file: nsvdis.tex
\documentclass[a4paper,11pt]{article}
\pdfoutput=1
\usepackage{jheppub}
\usepackage{graphicx,color}
\usepackage{amsmath,autobreak}
\usepackage{array}
\usepackage{ulem}
\newcolumntype{P}[1]{>{\centering\arraybackslash}p{#1}}
\allowdisplaybreaks
\RequirePackage{ifpdf}
\usepackage{amsmath}
\usepackage{mathtools}

\usepackage{jheppub}
\usepackage[final]{pdfpages}
\usepackage{ifpdf}
\usepackage{slashed}

\usepackage{hyperref}

\usepackage{color}
\usepackage{graphics}

\usepackage{etoolbox}
\usepackage{fixmath}

\usepackage{caption}
\usepackage{subcaption}
\usepackage{amsfonts}

\usepackage{multirow}
\usepackage{epstopdf}

\usepackage{relsize}
\usepackage{float}

\usepackage[table]{xcolor}
\usepackage{tabularx}

\def\arXiv{{\fontfamily{qcr}\selectfont arXiv}}

\usepackage{tikz}
\usetikzlibrary{positioning,arrows}
\usetikzlibrary{decorations.pathmorphing}
\usetikzlibrary{decorations.markings}
\usetikzlibrary{shapes.geometric}
\usepackage{endnotes}
\tikzset{
    vector/.style={decorate, decoration={snake}, draw},
    provector/.style={decorate, decoration={snake,amplitude=2.5pt}, draw},
    antivector/.style={decorate, decoration={snake,amplitude=-2.5pt}, draw},
    fermion/.style={draw=black,
      postaction={decorate},decoration={markings,mark=at position .55
        with {\arrow[draw=black]{>}}}},
    fermionbar/.style={draw=black, postaction={decorate},
                       decoration={markings,mark=at position .55 with {\arrow[draw=black]{<}}}},
    fermionnoarrow/.style={draw=black},
    gluon/.style={decorate, draw=black,decoration={coil,amplitude=4pt, segment length=6pt}},
    scalar/.style={dashed,draw=black,
      postaction={decorate},decoration={markings,mark=at position .55
        with {\arrow[draw=black]{>}}}},
    scalarbar/.style={dashed,draw=black,
      postaction={decorate},decoration={markings,mark=at position .55
        with {\arrow[draw=black]{<}}}},
    scalarnoarrow/.style={dashed,draw=black},
    electron/.style={draw=black,
      postaction={decorate},decoration={markings,mark=at position .55
        with {\arrow[draw=black]{>}}}},
    bigvector/.style={decorate, decoration={snake,amplitude=4pt}, draw},
}
\title{On next to soft threshold corrections to DIS and SIA processes}

\author{A.H. Ajjath, Pooja Mukherjee, V. Ravindran, Aparna Sankar and Surabhi Tiwari}
\emailAdd{
ajjathah@imsc.res.in,
poojamukherjee@imsc.res.in,
ravindra@imsc.res.in,
aparnas@imsc.res.in,
surabhit@imsc.res.in}

\affiliation{The Institute of Mathematical Sciences, HBNI, Taramani, Chennai 600113, India}

\preprint{IMSc/2020/11/04}

\abstract{
We study the perturbative structure of threshold enhanced logarithms 
in the coefficient functions of deep inelastic scattering (DIS) and semi-inclusive $e^+e^-$ annihilation (SIA)
processes and setup a framework to sum them up to all orders in perturbation theory.
Threshold logarithms show up as the distributions $((1-z)^{-1} \log^i(1-z))_+$ from
the soft plus virtual (SV) and as logarithms $\log^i(1-z)$ from next to SV (NSV) contributions.  
We use the Sudakov differential and the renormalisation group equations along with 
the factorisation properties of parton level cross sections to obtain 
the resummed result which predicts SV  as well as next to SV  
contributions to all orders in strong coupling constant.
In Mellin $N$ space, we resum the large logarithms of the form $\log^i(N)$ keeping $1/N$ corrections.
In particular, the towers of logarithms, each of the form  $a_s^n/N^\alpha \log^{2n-\alpha} (N), 
a_s^n/N^\alpha \log^{2n-1-\alpha}(N) \cdots $ etc for $\alpha =0,1$, are summed to all orders in $a_s$.}
\begin{document}

\allowdisplaybreaks[4]
\unitlength1cm
\maketitle
\flushbottom
\let\footnote=\endnote
\renewcommand*{\thefootnote}{\fnsymbol{footnote}}
\section{Introduction}
Radiative corrections to observables at high energy colliders are absolutely essential to
understand the underlying quantum dynamics of the scattering events.  There are a large number
of accurate measurements already available from these colliders and they provide ample opportunity
to investigate various theories that attempt to describe the physics.  Theoretical predictions
with unprecedented accuracy have already set stringent constraints on the parameters of the standard model (SM), also
for many beyond the SM (BSM) scenarios.  The observables that involve hadrons either in initial or in final state
receive large perturbative and non-perturbative quantum corrections from strong interaction which 
is described by quantum chromodynamics (QCD).
This is simply due to the strong coupling constant ($g_s$) which is big and due to a large number of scattering channels that
contribute.  At high energies, thanks to factorisation properties of certain hadronic observables,
which are infrared (IR) safe, the short distance perturbative part factorises from the
the non-perturbative one.  This allows one to reliably compute the perturbative quantum corrections in powers of strong coupling
constant $a_s = g_s^2/16 \pi^2$ in QCD.  The non-perturbative part of these IR safe observables are extracted in a process
independent way.  For example, the inclusive cross section of DIS of a lepton on 
a hadron factorises into perturbatively calculable coefficient functions (CF) and parton distribution functions (PDF)
that are non-perturbative in nature.  The CFs are computed in powers of $a_s$ using the parton level scattering
processes that contributes to the hadronic reaction.  The PDFs are nothing but the probability of finding a parton inside the hadron
during the scattering event, describing the long distance part of the hadronic events and hence can not be computed
using perturbative techniques.  
However, being process independent, they can be extracted from an experiment and can be used
for other experiments.   Within QCD, they are defined in terms of hadronic matrix elements of certain 
gauge invariant quantum field theoretical operators made up of quark, anti-quark and gluon field operators. These matrix elements satisfy the renormalisation group (RG) equations which go by the name Altarelli-Parisi (AP) or DGLAP evolution equations. The corresponding anomalous dimensions are called AP splitting functions.  Another example that is very similar to
DIS is the production of a hadron in  SIA  of $e^+ e^-$.  One finds that scattering cross section
in SIA also demonstrate factorisation of perturbatively calculable CFs and non-perturbative quantities called parton fragmentation
functions (PFF).  The CFs describe the production of a parton in the collision and the latter describes the
fragmentation of the produced parton into a hadron.  PFFs also satisfy AP or DGLAP equations with the corresponding
AP splitting functions.  Denoting $\sigma_I$ generically for the inclusive cross section for DIS ($I=DIS$) and SIA ($I=SIA$),
the factorisation at high energies implies
\begin{eqnarray}
\label{genSIGI}  
\sigma_I(Q^2,x_I) = \sigma_{I}^{(0)}(Q^2,\mu_R^2) \sum_{a=q,\overline q,g} \int dx f_a(\mu_F^2,x) \Delta_{I,a}(Q^2,\mu_R^2,\mu_F^2,z_I)
+ {\cal O}(1/Q^2)\,.
\end{eqnarray}
Since, we study these observables in the large $Q^2$ region, we drop the power suppressed contributions denoted by ${\cal O}(1/Q^2)$ in
the above formula and consider only the first term for rest of our study. 
$\sigma_{I}^{(0)}$ is the born level cross section and $\mu_R$ is the ultraviolet renormalisation scale,  
$f_a$ denotes PDF for $I=DIS$ and PFF for $I=SIA$. The PDFs depend on the partonic momentum fraction $x$
carried away from the hadron in DIS and PFFs depend on the hadronic momentum fraction that the hadron 
carries away from the parton.  The scale $\mu_F$ is called the factorisation
scale which separates perturbative and non-perturbative regions.  The sum is over all the partons namely the quarks and anti-quarks of
all flavours and the gluons.  The scale $Q^2$ is the hard scale in the problem.  For DIS, it is defined by $Q^2 =-q^2$, $q$ being
the momentum that is transferred from the incoming lepton to the target hadron.  The corresponding scaling variable 
$x_{DIS} = Q^2/2 P \cdot q$ where $P$ is the momentum of the target hadron.  Similarly, for SIA, $Q^2 = q^2$ with $q$ being the sum of momenta
of incoming leptons and $x_{SIA}= 2 P\cdot q/Q^2$, with $P$ being the momentum of hadron that fragments from the parton.  
The parton level scaling variables are  $z_{DIS} = Q^2/2 p\cdot q$ and $z_{SIA} = 2 p \cdot q/Q^2$.  Here, $p$ is the momentum of
the parton in the respective scattering processes.  In the rest of our paper, we drop $I$ in the argument of $\Delta_{I,a}$.

Perturbative QCD provides framework to compute $\Delta_{I,a}$ in powers of $a_s$:
\begin{eqnarray}
\label{genDELI}
\Delta_{I,a}(Q^2,\mu_R^2,\mu_F^2,z) = \sum_{i=0}^\infty a_s^i(\mu_R^2) \Delta_{I,a}^{(i)}(Q^2,\mu_R^2,\mu_F^2,z)\,,
\end{eqnarray}
where $\Delta_{I,a}^{(i)}$ at every order gets contribution from the parton level scattering processes.
Beyond leading order in perturbation theory, at the intermediate stages they contain ultraviolet (UV), soft and collinear
divergences.  The UV divergences go away when renormalisation of coupling, masses and fields are
performed.   The soft and collinear divergences are collectively called infrared (IR) divergences.
The soft divergences come from zero momentum gluons in the loops of virtual contributions and real gluons
in the gluon emission processes.  The massless or light partons are responsible for collinear divergences.    
Thanks to KLN theorem \cite{Kinoshita:1962ur,Lee:1964is}, soft and collinear divergences go away when the degenerate states that are
responsible are summed at the partonic level.  However, for DIS, the sum over degenerate partonic initial states 
are summed through convoluting them with bare PDFs.  In practice, the initial state collinear divergences
are factored out from the partonic subprocesses and then absorbed into the bare PDFs.  This is called
mass factorisation.  Similar thing happens for the SIA where the final state collinear singularities are absorbed into 
bare PFFs to get IR safe observable.  The factorisation scale quantifies the arbitrariness involved in the mass factorisation.
In practice, both UV and IR divergences are regulated in dimensional regularisation by working in complex space time
dimension $n = 4 + \epsilon$.  The divergences show up as poles in $\epsilon$.  The UV renormalisation and mass 
factorisation are done in modified minimal substraction $\overline {MS}$ scheme consistently.
The inclusive cross sections for DIS 
are known to third order in QCD,  see \cite{Vermaseren:2005qc,Soar:2009yh,Ablinger:2014vwa}.
%and \cite{Anastasiou:2015ema,
%Mistlberger:2018etf,Duhr:2019kwi} respectively and for invariant mass distribution up to third order in QCD 
%see \cite{Hamberg:1990np,Harlander:2002wh,Duhr:2020seh}, for complete list see \cite{Georgi:1977gs,Graudenz:1992pv,Djouadi:1991tka,Spira:1995rr,Catani:2001ic,Harlander:2001is,Anastasiou:2002yz,Harlander:2002wh,Catani:2003zt,Ravindran:2003um,Moch:2005ky,Ravindran:2006cg,deFlorian:2012za,Bonvini:2014jma,deFlorian:2014vta,Anastasiou:2014vaa,Li:2014afw,Anastasiou:2015yha,Anastasiou:2015vya,Das:2020adl} for Higgs production in gluon fusion 
%and \cite{Altarelli:1978id,Altarelli:1979ub,Matsuura:1987wt,Matsuura:1988nd,Matsuura:1988sm,Matsuura:1990ba,Hamberg:1990np,vanNeerven:1991gh,Harlander:2002wh,Moch:2005ky,Ravindran:2006cg,deFlorian:2012za,Ahmed:2014cla,Catani:2014uta,Li:2014afw,Duhr:2020seh} for Drell-Yan production.

In CFs, the energy scales $Q^2,\mu_R^2$ and $\mu_F^2$ appear as logarithms and the  partonic scaling variable shows up through 
$\delta(1-z)$, plus distributions ${\cal D}_j = (\log^j(1-z)/(1-z))_+$ and regular functions of $z$:
\begin{eqnarray}  
\label{DeltaI}
\Delta_{I,a}(z) = \Delta_{I,a,\delta} \delta(1-z) + \sum_{j=0}^\infty \Delta_{I,a,{\cal D}_j} \left( {\log^j(1-z) \over 1-z } \right)_+
+ \Delta_{I,a,R}(z)\,,
\end{eqnarray}
where we have suppressed the scales $Q^2,\mu_R^2$ and $\mu_F^2$ in the arguments of 
$\Delta_{I,a}$ and $\Delta_{I,a,Z},Z=\delta,{\cal D}_j,R$ on both sides.
Large number of perturbative results provide opportunity to understand the universal structure of IR divergences.  
For example, the IR structure of multi-leg amplitudes in QCD is well understood beyond two loop
level \cite{Catani:1998bh,Becher:2009cu,Becher:2009qa,
Gardi:2009qi,Catani:1998bh} (see \cite{Ajjath:2019vmf,H:2019nsw} for a QFT with mixed gauge groups).
In addition, we have large number of results for the inclusive cross sections that can shed light on
the structure of $\Delta_{I,a,Z}$, see 
\cite{Anastasiou:2015vya, Mistlberger:2018etf,Duhr:2019kwi} for Higgs production 
and for invariant mass distribution of a pair of lepton in hadron colliders up to third order in QCD
see \cite{Hamberg:1990np,Harlander:2002wh,Duhr:2020seh}, for complete list see \cite{Georgi:1977gs,Graudenz:1992pv,Djouadi:1991tka,Spira:1995rr,Catani:2001ic,Harlander:2001is,Anastasiou:2002yz,Harlander:2002wh,Catani:2003zt,Ravindran:2003um,Moch:2005ky,Ravindran:2006cg,deFlorian:2012za,Bonvini:2014jma,deFlorian:2014vta,Anastasiou:2014vaa,Li:2014afw,Anastasiou:2015yha,Anastasiou:2015vya,Das:2020adl} for Higgs production in gluon fusion
and \cite{Altarelli:1978id,Altarelli:1979ub,Matsuura:1987wt,Matsuura:1988nd,Matsuura:1988sm,Matsuura:1990ba,Hamberg:1990np,vanNeerven:1991gh,Harlander:2002wh,Moch:2005ky,Ravindran:2006cg,deFlorian:2012za,Ahmed:2014cla,Catani:2014uta,Li:2014afw,Duhr:2020seh} for Drell-Yan production.

The distributions $\delta(1-z)$ and ${\cal D}_j(z)$ result from the soft and collinear regions of the virtual and real emission diagrams.
In the region where a scattering event involves infinite number of
soft gluons each carrying almost zero momentum, 
the logarithms of the form $\log^i(1-z)/(1-z)$
contribute to $\Delta_{I,a}$.  This can happen in real emission scattering processes.
These contributions are ill defined in 4 space-time dimensions in the limit $z \rightarrow 1$. 
The inclusion of these contribution
gives the distributions ${\cal D}_i(z)$ and $\delta(1-z)$.
The distributions that are present in $\Delta_{I,a}$ are called soft plus virtual (SV) contributions.
SV results are available for many observables at colliders up to third order in QCD, 
see \cite{Moch:2005ky,Ravindran:2005vv,Ravindran:2006cg,deFlorian:2012za,Ahmed:2014cha, Kumar:2014uwa,
Ahmed:2014cla,Catani:2014uta,Li:2014bfa}.
When the distributions are convoluted with PDFs or PFFs  to obtain hadronic cross section, one finds that they not only dominate over
other contributions but also are large at every order.  Hence, they can 
spoil the reliability of the predictions from the truncated series.
The resolution to this problem was successfully achieved in seminal works by 
Sterman \cite{Sterman:1986aj} and Catani and Trentedue \cite{Catani:1989ne}
through reorganisation of the perturbative series.   It goes under the name threshold resummation, see also
\cite{Catani:1996yz,Moch:2005ba,Bonvini:2012an,Bonvini:2014joa,Bonvini:2014tea,Bonvini:2016frm} for Higgs production in gluon fusion,
\cite{Bonvini:2016fgf,H:2019dcl} for bottom quark annihilation, for DY \cite{Moch:2005ba,Bonvini:2010ny,Bonvini:2012sh,
H.:2020ecd,Catani:2014uta} and for DIS and SIA of $e^+ e^-$ \cite{Cacciari:2001cw}.  In Mellin space, the conjugate variable to $z$ is $N$ and the convolutions
become normal products.  Hence, the resummation is conveniently done in Mellin space.
In the Mellin space, the threshold limit $z \rightarrow 1$ corresponds to large $N$.  The large logarithms
of $N$ at every order combined with the strong coupling constant can give order one contribution.  Hence,
the truncation of the series based on series expansion in $a_s$ is not allowed.  However, thanks to 
factorisation properties, universality of IR contributions and renormalisation group invariance,
we can systematically resum the order one terms, in particular, terms of the form $a_s(\mu_R^2) \beta_0 \log(N)$ to all orders    
in perturbation theory.  Defining $a_s(\mu_R^2) \beta_0 \log(N) = \omega$, and treating $\omega$ to order 1, following,
\cite{Sterman:1986aj,Catani:1989ne}, we can organize 
\begin{eqnarray}
\label{resumgen}
\lim_{N\rightarrow \infty} \log \Delta_{I,a,N} = 
 \log \tilde g^{I}_0(a_s(\mu_R^2))+
\log(N) g^{I}_1(\omega) + \sum_{i=0}^\infty a_s^i(\mu_R^2) g^{I}_{i+2}(\omega) \,,
\end{eqnarray} 
where $\tilde g^{I}_0(a_s(\mu_R^2))$ is $N$ independent.
Inclusion of  successive terms in Eq.(\ref{resumgen}) predicts the leading logarithms (LL), 
next to leading (NLL) etc logarithms to all orders in $a_s$.  
The functions $g^{I}_i(\omega)$ depend on universal  
IR anomalous dimensions and  $\tilde g^I_0$ depend on the hard process. 
For DIS, invariant mass distribution of lepton pairs in DY, Higgs boson productions in various 
channels, results for the resummation of threshold logarithms in $N$ space up to third order, namely
next to next to next to leading logarithmic (N$^3$LL) accuracy, are available\cite{Catani:2014uta,Moch:2005ba,H.:2020ecd,H:2019dcl}. 

The resummed predictions played an important role to understand the experimental data in the threshold regions.
However, the sub leading logarithms that are present in the regular part $\Delta_{I,a,R}(z)$ can not
be ignored. We expand $\Delta_{I,a,R}(z)$ around $z=1$ to obtain  
\begin{eqnarray}
\label{DeltaR}
\Delta_{I,a,R}(z) = \sum_{k=0}^\infty \Delta_{I,a,L}^{(k)} \log^k(1-z) + {\cal O}(1-z)
\end{eqnarray}
where the logarithms of the form $\log^k(1-z),k=0,1,\cdot \cdot \cdot$ do contribute
significantly at every order in perturbation theory near the threshold.   
We call them by next to SV (NSV) contributions.  There have been several studies to understand
the structure of NSV terms in the hadronic observables so that one can find whether 
the NSV terms can be systematically resummed to all orders  like the way the SV terms are resummed.
There have been several attempts to achieve this task.  
A remarkable development was made by Moch and Vogt
in \cite{Moch:2009hr} (and \cite{deFlorian:2014vta,Das:2020adl})
using the second order results for DIS, semi-inclusive $e^+ e^-$ annihilation and Drell-Yan production of
a pair of leptons in hadron collisions,
and the  physical evolution kernels  to find the enhancement of a single-logarithms at large $z$ to all orders in $1-z$.
The physical evolution kernel was exploited earlier in the work by \cite{Grunberg:2009yi}.
It was found that the structure of corresponding
leading $\log(1-z)$ terms in the kernel can be constrained \cite{Moch:2009hr} allowing them  to predict
certain next to SV logarithms at  higher orders in $a_s$. 
The next to SV corrections to various inclusive processes were studied in a series of papers \cite{Laenen:2008ux,Laenen:2010kp,Laenen:2010uz,Bonocore:2014wua,Bonocore:2015esa,Beneke:2019oqx,Beneke:2019mua,Bonocore:2016awd,DelDuca:2017twk} and 
much progress have been made which lead to better understanding of the underlying physics.
Recently some of us
have studied inclusive production of pair of leptons in Drell-Yan process and of a Higgs boson in gluon fusion as well as
in bottom quark annihilation in an attempt to resum these NSV terms to all orders\cite{Ajjath:2020ulr}.  We used
factorisation properties and renormalisation group invariance along with the certain universal structure
of real and virtual contributions using Sudakov K+G equation to achieve this task.   In this paper, 
we extend this to DIS and SIA to provide an all order result both in $z$ space and in $N$ space.  

% section 2

\section{Next to SV in $z$ space}
We begin with the unpolarised inclusive deep-inelastic lepton-nucleon scattering:
\begin{eqnarray}
\label{procdis}
l(k) + H(P) \rightarrow l(k') + X(P_X)\,,
\end{eqnarray}
where the incoming and scattered leptons ($l$) carry the momenta $k$ and $k'$ respectively,  $H$ is the target hadron 
with the momentum $P$ and the $X$ is the set of inclusive final states with total momentum $P_X$.  
If we restrict to only photon exchange in the
scattering, the inclusive cross section can be expressed in terms of
two structure functions (SF) namely $F_1(Q^2,x)$ and $F_2(Q^2,x)$.  The SFs are scalar functions and they parametrise
the hadronic tensor $W_{\mu \nu}(Q^2,x)$ which carry the information of hadronic part of the DIS cross section.  
The tensor $W_{\mu \nu}$ is given by   
\begin{eqnarray}
\label{Wmunu}
W_{\mu \nu}(Q^2,x) = \left( {q_\mu q_\nu \over q^2} - g_{\mu \nu} \right) F_1(Q^2,x)
- {1 \over 2 x q^2} \Big(q_\mu + 2 x P_\mu\Big)\Big(q_\nu + 2 x P_\nu\Big)  F_2(Q^2,x)\,,
\end{eqnarray}
where $q=k'-k$, $Q^2=-q^2$ and the scaling variable, also called Bj\"orken $x$ is defined by $x=Q^2/2 P\cdot q$.     
The hadronic tensor is related to Fourier transform of 
commutator of two electromagnetic currents sandwiched between the hadronic states.  Due to the non-perturbative
nature of the hadronic states, 
the structure functions are not computable in perturbation theory.  However, in the Bj\"orken limit,
thanks to operator product expansion, the hadronic tensor factorises into perturbatively calculable 
Wilson coefficients and non-perturbative composite operators sandwiched between hadronic states.  
Defining the Mellin moment of $F_i(Q^2,x)$ by 
\begin{eqnarray}
\label{mellin}
F_{J,N}(Q^2) = \int_0^1 dx x^{N-1} {F_J(Q^2,x) \over x}\,, \quad \quad J = 2,L
\end{eqnarray}
with $F_L = F_2 -2 x F_1$ and 
computing them in the Bj\"oken limit, namely $Q^2 \rightarrow \infty,P\cdot q \rightarrow \infty$ keeping
$x=Q^2/2 P\cdot q$ fixed, one finds 
\begin{eqnarray}
\label{ope}
F_{J,N}(Q^2) = \sum_{a=ns,q,g} C_{J,a,N}(Q^2,\mu_F^2) {\cal A}_{a,N}(\mu_F^2)\,,
\end{eqnarray}
where $ns$ denotes the non-singlet combination of quark operators, which does not mix with the 
gluonic operator under 
UV renormalisation and the indices $q$ and $g$ correspond to those operators which mix among themselves.
The matrix element of local operators denoted by ${\cal A}_{a,N}$ are not calculable using perturbative methods.
However, their evolution in terms of the scale $\mu_F$ is controlled by the perturbatively calculable AP splitting functions
through AP evolution equations.  The Wilson coefficients $C_{J,a,N}$ are computed in powers of strong coupling constant.   

In QCD improved parton model, one can relate the local operators ${\cal A}_{a,N}$ to Mellin moments of
appropriate combinations of PDFs and the Wilson coefficients $C_{J,a,N}$ to parton level coefficient functions (CFs). 
The Wilson coefficients, equivalently CFs can be computed, within the framework of perturbative QCD, 
order by order in strong coupling constant using parton level subprocesses. The contributions,
beyond the leading order, contain UV, soft and collinear divergences.  If we regulate them 
in dimensional regularisation, the UV divergences 
arise as poles in $\epsilon$ and are removed in modified minimal subtraction ($\overline {MS}$) scheme. 
As we discussed in the introduction, the soft and collinear divergences resulting from final 
state partons cancel independently 
after summing up the contributions from all  
possible degenerate states. However, the collinear divergences arising 
from the initial state
light partons remain. Those are removed at the hadronic level through a procedure called mass factorisation. 

In the following, we consider the Wilson coefficients equivalent to the CFs that contribute to a generic DIS scattering process.  
We denote them by $\Delta_c$ where the index $c=q,\overline q, g$.   
The factorisation allows us to relate the CFs, $\Delta_{c}$ and  the parton level subprocesses 
through the mass factorisation given as
\textcolor{black}{
\begin{eqnarray}
\label{MassFact}
{1 \over z} {\hat \sigma_{c} (Q^2,z,\epsilon) } &=& \sigma^{(0)}(\mu_R^2) \Gamma_{c'c}(z,\mu_F^2,\epsilon) \otimes \left({1 \over z} 
	C_{c'}(Q^2,\mu_R^2,\mu_F^2,z,\epsilon)\right) \,.
\end{eqnarray}}
In Eq.(\ref{MassFact}), 
$\hat \sigma_{c}(Q^2,z,\epsilon)/z$ is the appropriate UV finite parton level cross section computed in space time 
dimension $n=4+\epsilon$.
The scaling variable $z$ is given by $z=Q^2/2 p\cdot q$, where $p$ is the momentum of the incoming parton in the
scattering event.
The function $\Gamma_{c' c}$ is the Altarelli-Parisi \textcolor{black}{(AP) \cite{Altarelli:1977zs}} kernel which 
contains the collinear divergences of $\hat \sigma_c$ in ${\overline {MS}}$ scheme. 
As in \cite{Ajjath:2020ulr}, we limit ourselves to SV+NSV contributions to CFs, which means that we 
drop those terms in $C_c$ that vanish when $z\rightarrow 1$ and call the resulting ones by $\Delta_c$.  
For the quark/anti-quark initiated processes in DIS with photon exchange, gluon initiated one with 
Higgs boson exchange, the infrared singluar partonic cross sections can be factorised 
into squares of UV renormalisation constant, $Z_{UV,c}^2$, of form factor (FF), $|\hat F_c|^2$ and a function ${\cal S}_J^c$ 
that is sensitive to real radiations. This is always possible as $Z_{UV,c}^2$ and $|\hat{F}_c|^2$ are simply proportional to
$\delta(1-z)$ and can be factored out from these partonic channels. That is,
\begin{eqnarray}
\label{normS}
z^{-1}\hat \sigma_c (Q^2,z,\epsilon) & =&	\sigma^{(0)}(\mu_R^2) \left(Z_{UV,c}(\hat{a}_s,\mu_R^2,\mu^2,\epsilon) \right)^{2}
|\hat F_{c}(\hat a_s,\mu^2,Q^2,\epsilon)|^{2}
\nonumber\\
%&\qquad\qquad~~~
&&\times \delta(1-z) 
 \otimes {\cal S}_J^c\left(\hat a_s,\mu^2,Q^2,z,\epsilon\right)
%\nonumber \\
%&& 
%\otimes {\cal C} \exp\left(2 {\mathrm \Phi}_J^c(\hat a_s,\mu^2,Q^2,z,\epsilon)\right) \,.
%\otimes \Gamma^{-1}_{cc}(\hat a_s,\mu^2,\mu_F^2,z,\epsilon).
\end{eqnarray}
As it will be shown in the following, the function
${\cal S}_J^c$ satisfies a differential equation which admits a solution namely the convoluted exponential of $\Phi_c$.
That is, 
\begin{eqnarray}
\label{calS}
{\cal S}_J^c = {\cal C}\exp\left(2 {\mathrm \Phi}_J^c(\hat a_s,\mu^2,Q^2,z,\epsilon)\right) \,.
\end{eqnarray} 
Substituting for $\hat \sigma_c$ from (\ref{normS}) in terms of ${\mathrm \Phi}_J^c$ in (\ref{MassFact}), we obtain
\begin{eqnarray}
\label{MasterF}
 \Delta_{c}(Q^2,\mu_R^2,\mu_F^2,z) & =& \mathcal{C}\exp \bigg( \Psi_J^c\big(Q^2,\mu_R^2,\mu_F^2,z,\epsilon\big)\bigg)\bigg |_{\epsilon=0} \,,
\end{eqnarray}
where the function $\Psi_J^c$ is given by
\begin{eqnarray}\label{Psi}
    \Psi_J^c\big(Q^2,\mu_R^2,\mu_F^2,z,\epsilon\big) = &\Bigg( \ln \bigg( Z_{UV,c}\big(\hat{a}_s,\mu^2,\mu_R^2,\epsilon\big)\bigg)^2 +   \ln \big| \hat{F}_{c}\big(\hat{a}_s,\mu^2,Q^2,\epsilon\big)\big|^2\Bigg) \delta\big(1-z\big)\nonumber \\
    &+2 \mathrm{\Phi}_J^c\big(\hat{a}_s,\mu^2,Q^2,z,\epsilon\big) - \mathcal{C} \ln \Gamma_{cc}\big(\hat{a}_s,\mu^2,\mu_F^2,z,\epsilon\big) \,.
\end{eqnarray}
The \textcolor{black}{symbol $\otimes$ represents the Mellin convolution}. 
The operation of $\mathcal{C}$ on any given function is defined in Eq.(2) of \cite{Ravindran:2005vv}. 
In this expression, $c = q$ (quark/antiquark) for photon-exchange DIS,
and $c = g$ (gluon) for Higgs-exchange DIS.
Though the constituents of $\Psi_J^c$ contains UV and IR divergent terms, the sum of all these terms is finite 
and is regular in the variable $\epsilon$.  It contains the distributions such as
$\delta(1-z)$, ${\cal D}_j(z)$ and the logarithms of the form $\log^i(1-z), i=0,1,\cdots$.
In Eq.(\ref{Psi}), the overall renormalisation constant for DIS via the photon exchange is one to all orders in QCD. 
For DIS via the Higgs boson exchange, $Z_{UV,c}$ is equivalent to that of Higgs-gluon 
effective operator\cite{Ravindran:2005vv}.

The AP kernels that remove collinear divergences from the parton level cross sections are 
solutions to AP evolution equation (see Eq. (2.11) in 
\citep{Ajjath:2020ulr}) which are controlled by AP splitting functions $P_{ab}(\mu_F^2,z)$.
They contain convolutions of AP spitting functions.
In the above equation, we have kept only diagonal part of AP kernel $\Gamma_{ab}$ and 
dropped the non-diagonal AP kernels.
We explain the reason below.
Consider $\Delta_q$ in photon-exchange DIS.  It gets contributions from three different terms 
namely $\hat\sigma_{q } \otimes \Gamma_{qq}$, $\hat\sigma_{\overline q } \otimes \Gamma_{\overline q q}$ 
and  $\hat\sigma_{g } \otimes \Gamma_{g q}$. 
The non-diagonal AP kernels and $\hat \sigma_g$ contain only NSV and/or beyond NSV terms.
Upon convolution, the terms  $\hat\sigma_{\overline q } \otimes \Gamma_{\overline q q}$
and  $\hat\sigma_{g } \otimes \Gamma_{gq}$ will give only beyond NSV terms.
In addition, only diagonal parts of splitting functions $P_{ab}(z, \mu_F^2)$ in $\Gamma_{ab}(z,\mu_F^2, \epsilon)$ 
need to be kept as the contributions from convolutions of two or more non-diagonal splitting functions give 
only beyond NSV terms.  
The diagonal  $P_{c c}\big(z,\mu_F^2\big)$ are expanded
around $z=1$ and all those terms that do not contribute to SV+NSV are eliminated.  The diagonal AP splitting functions
near $z=1$ take the following form:
\begin{eqnarray}
        P_{cc}\big(z,a_s(\mu_F^2)\big) &=& 2 B^c(a_s(\mu_F^2)) \delta(1-z) + 2  P^{\prime}_{cc}\big(z,a_s(\mu_F^2)\big)\,,
\end{eqnarray}
where,
\begin{eqnarray}
        P^{\prime}_{cc}\big(z,a_s(\mu_F^2)\big) &=&  \Bigg[ A^c(a_s(\mu_F^2)) {\cal D}_0(z)
%\nonumber\\&&
                      + C^c(a_s(\mu_F^2)) \log(1-z) + D^c(a_s(\mu_F^2)) \Bigg].
\end{eqnarray}

%\begin{eqnarray}
%        P_{cc}\big(z,a_s(\mu_F^2)\big) &=& 2  \Bigg[ B^c(a_s(\mu_F^2)) \delta(1-z) + A^c(a_s(\mu_F^2)) {\cal D}_0(z)
%\nonumber\\&&
%                      + C^c(a_s(\mu_F^2)) \log(1-z) + D^c(a_s(\mu_F^2)) \Bigg]  \,.
%\end{eqnarray}
The constants $C^c$ and $D^c$ can be obtained from the
the splitting functions $P_{c c}$ which are known to three loops in QCD \cite{Moch:2004pa,Vogt:2004mw}
(see \cite{GonzalezArroyo:1979df,Curci:1980uw,Furmanski:1980cm,Hamberg:1991qt,Ellis:1996nn,Moch:2004pa,Vogt:2004mw,Soar:2009yh,Ablinger:2017tan,Moch:2017uml} for the lower order ones).

For the DIS with the photon exchange, the interaction of
virtual photon from the lepton with the target hadron is through a vector current.  Hence, the FF that
contributes to the inclusive cross section is the square of the quark matrix element for the vector current.
Vector current being conserved does not get any overall UV renormalisation and hence $Z_{UV,c}$ is identity.  
If the exchange particle is the scalar Higgs boson and its interaction with the hadron is through an operator
which is not conserved, then $Z_{UV,c}$ will be non-zero.  For example, in order to compute singlet splitting
functions, one resorts to scattering of a scalar Higgs boson on a gluon target and the interaction between
them is governed by effective operator $G_{\mu \nu}^a G^{\mu \nu, a} \phi$.  Here $G_{\mu\nu}^a$ is the gluon
field strength operator and $\phi$ is the Higgs boson field.  The coupling of Higgs boson and the gluon through
this composite operator requires addition overall renormalisation and hence $Z_{UV,c}$ \cite{Chetyrkin:2005ia} is included in 
Eq.(\ref{Psi}).  
FFs in general are computable in regularised QCD perturbation theory in powers of strong coupling constant.  
FFs are known in QCD up to third order in perturbation theory,
\cite{vanNeerven:1985xr,Harlander:2000mg,Ravindran:2004mb,Moch:2005tm,Gehrmann:2005pd,Baikov:2009bg,Gehrmann:2010ue,Gehrmann:2014vha,vonManteuffel:2016xki,Henn:2016men,Henn:2019rmi,vonManteuffel:2020vjv,Gehrmann:2010tu}.
Both UV and IR divergences appear as poles in $\epsilon$ and they  
demonstrate rich IR structure, and satisfy differential equations such as 
RG equation, $\mu_R^2\dfrac{ d }{d\mu_R^2}\hat F_c = 0$ and 
Sudakov differential equation \cite{Sudakov:1954sw,Sen:1981sd,Collins:1989bt,Magnea:1990zb,Magnea:2000ss,Sterman:2002qn,Moch:2005id,Ravindran:2005vv}. The latter is called K+G equation. It is used to study their IR structure of FFs 
in terms of IR
anomalous dimensions such as cusp $A^c$,collinear $B^c$ and soft $f^c$ anomalous dimensions.
The perturbative structure of FFs provides valuable information of the underlying quantum field theory and it was exploited
to understand the structure of multi-leg on-shell amplitudes in QCD \cite{Catani:1998bh,Becher:2009cu,Becher:2009qa,
Gardi:2009qi,Catani:1998bh} (see \cite{Ajjath:2019vmf,H:2019nsw} for a QFT with mixed gauge groups) 
and they are found to be helpful to  understand the IR structure of real emission processes \cite{Ravindran:2005vv,Ravindran:2006cg,Ahmed:2014cha, 
Kumar:2014uwa,Ahmed:2014cla}. 
 
In \cite{Ravindran:2005vv,Ravindran:2006cg}, using the K+G structure of FF and the
finiteness of inclusive cross sections, it was shown that the soft distribution functions \textcolor{black}{${\cal S}_J^c$, equivalently, ${\mathrm \Phi}_J^c$} in Drell-Yan production
of lepton pairs and production of Higgs boson in gluon fusion in hadron colliders and soft plus jet function in
DIS processes were shown to satisfy K+G type differential equations.  The infrared structure of these
functions can be understood in terms of the IR anomalous dimensions.  In particular, the threshold logarithms that
contribute in the soft and collinear regions of the real emission processes are contained in these soft functions.  The
universal nature of these contributions are due to the IR anomalous dimensions.   The fact that these contributions
exponentiate, owing to the K+G differential equation that they satisfy, the all order predictions as well as
the resummation of threshold effects are possible.  In the present case, our task is to find a suitable K+G equation which
can capture not only SV contributions but also NSV contributions.

{Using the fact that the function ${\cal S}_J^c$ given in (\ref{normS}) can be factorised from the rest of the contributions and that the FF satisfies K+G equation, we can easily show that ${\cal S}_J^c$ also satisfies a K+G type differential equation.  Note that ${\cal S}_J^c$ captures both soft and next to soft contributions.   Since the K+G equation corresponding to ${\cal S}_J^c$ admits a solution of convoluted exponential form, we have expressed ${\cal S}_J^c = {\cal C} \exp\left(\mathrm \Phi_J^c\right)$ as given in (\ref{calS}), where the real emission contributions, normalised by $|\hat F^c|^2$ and $Z^2_{UV,c}$ are encapsulated in the function $\mathrm \Phi_J^c$.  Here, the exponential form of the real emission contributions holds true for both SV and NSV cases as the factorisation and the K+G differential equation are valid for all $z$.  We can use the finiteness of the coefficient function, $\Delta_c$ to determine ${\cal S}_J^c$ order by order in perturbation theory.}
In summary, we find that ${\cal S}_J^c$, equivalently 
$\mathrm \Phi_J^c$ satisfies K+G type equation
with the kernels $\overline K_J^c$ and $\overline G_J^c$ which contain right IR divergences and the finite terms respectively:
\begin{align}\label{KGphi}
Q^2\frac{d}{dQ^2}{\mathrm{\Phi_J}}^c = \frac{1}{2} \Big[ \overline{K}_J^c \Big(& \hat{a}_s, \frac{\mu_R^2}{\mu^2},\epsilon,z \Big) 
+ \overline{G}_J^c \Big( \hat{a}_s,\frac{Q^2}{\mu_R^2},\frac{\mu_R^2}{\mu^2},\epsilon,z \Big) \Big] \,.
\end{align}
Note that both $\overline K_J^c$ and $\overline G_J^c$ that control the evolution of $\mathrm \Phi_J^c$ are dependent on $z$. In addition,
following the structure of K+G equation for the FF, we keep all the IR divergent terms in $\overline K^c_J$ and move the entire $Q^2$
dependence along with IR finite terms  to $\overline G_J^c$. 
This is possible to all orders thanks to the factorisation property of real emission contributions. 
Following \cite{Ravindran:2005vv,Ravindran:2006cg}, we find the solution to 
Eq. (\ref{KGphi}).  Expanding both $\overline K_J^c$ and $\overline G_J^c$ in powers of bare coupling constant $\hat a_s$
and integrating over $Q^2$, we find 
\begin{align}\label{PhiSV1}
\mathrm{\Phi_J}^{c}(\hat{a}_s, Q^2,\mu^2,z, &\epsilon) = 
\sum_{i=1}^\infty\hat{a}_s^i\Big(\frac{Q^2(1-z)}{\mu^2 z}\Big)^{i\frac{\epsilon}{2}} S_{\epsilon}^{i}
\Big(\frac{i\epsilon}{2 (1-z)}\Big)\hat{\phi}_{c}^{(i)}(z,\epsilon)\,.
\end{align}
Few comments on the solution are in order.  The solution satisfies RG equation, namely $\mu_R^2  \dfrac{ d}{d\mu_R^2}\mathrm \Phi_J^c =0$ which
organises the perturbative expansion in such a way that after UV renormalisation, $\mathrm \Phi_J^c$ is free of UV divergences.  
In addition, it controls
the structure of logarithms of $Q^2$ through the term $Q^{i \epsilon}$.  Hence, $\mathrm \Phi_J^c$ contains only IR divergences 
and they are organised in such a way that they cancel against those from the FF and AP kernel.   
The factor $((1-z)/z)^{\epsilon/2}$
is inspired from the two body phase of the next to leading order DIS scattering and the term $1/(1-z)$ results form
the dominant contribution of the square of the parton level cross section in the limit $z \rightarrow 1$.  
The regular function denoted by $\hat \phi^{(i)}_c(z,\epsilon)$ determines the SV as well as NSV terms systematically
when it is expanded around $z=1$.
We determine the entire IR divergences in $\overline K_J^c$ 
from the those of $K_J^c$ of FF and of the SV part of the AP kernel 
by demanding IR finiteness of the SV part of the $\Delta_c$.  The remaining collinear divergences present in the AP kernel, 
which are sensitive to NSV terms, determine $\overline G_J^c$ with the condition of IR finiteness of $\Delta_c$ implied.  

For convenience, we decompose $\mathrm{\Phi}_J^c$ as $\mathrm{\Phi}_J^c$ = $\mathrm{\Phi}_{J,A}^{c}$ + $\mathrm{\Phi}_{J,B}^{c}$ 
in such a way that $\mathrm{\Phi}_{J,A}^{c}$ contains only SV terms $i.e$ all the distributions ${\cal D}_j$ and $\delta(1-z)$ and $\mathrm{\Phi}_{J,B}^{c}$ contains NSV terms namely $\log^k(1-z), k=0,1,\cdots$ in the limit $z \rightarrow 1$. 
An all order solution for $\mathrm{\Phi}_{J,A}^{c}$ in powers of  $\hat a_s$ in dimensional regularisation
is given in \cite{Ravindran:2005vv} and we reproduce here for completeness: 
\begin{align}\label{PhiSV}
\mathrm{\Phi}_{J,A}^{c}(\hat{a}_s,
Q^2,\mu^2, &\epsilon,z) = \sum_{i=1}^\infty\hat{a}_s^i\Big(\frac{Q^2(1-z)}{\mu^2}\Big)^{i\frac{\epsilon}{2}}
 S_{\epsilon}^{i}
\Big(\frac{i\epsilon}{2(1-z)}\Big)\hat{\phi}_{SV}^{c (i)}(\epsilon) \,, 
\end{align} 
where,
\begin{eqnarray}
\hat{\phi}_{SV}^{c(i)}(\epsilon) = \frac{1}{i\epsilon}\Big[ \overline{K}_J^{c(i)}(\epsilon) + \overline{G}_{J,SV}^{c(i)}(\epsilon) \Big]\,.
\end{eqnarray} 
The constants $\overline{K}_J^{c(i)}(\epsilon)$ and $\overline{G}_{J,SV}^{c(i)}(\epsilon) $ are given in Eq.(35)  and Eq.(37) of \cite{Ravindran:2006cg} respectively and they are known up to third order in perturbation theory
\cite{Ravindran:2005vv,Ravindran:2006cg,Bauer:2003pi,Bosch:2004th,Becher:2009th,Becher:2006qw,Becher:2010pd,Bruser:2018rad,Banerjee:2018ozf}.  The distributions in $\mathrm \Phi_J^c$ are related to Jet functions which 
are building blocks in   
Soft-Collinear effective theory (SCET) \cite{Bauer:2000ew,Bauer:2000yr,Bauer:2001ct,Bauer:2001yt,Bauer:2002nz,Beneke:2002ph} which captures the physics of soft and collinear dynamics 
of high energy scattering processes through the soft and jet functions. 
The jet functions describe the propagation of collinear partons inside jets. 
In SCET, the quark and gluon jet functions have been computed to higher orders in perturbation theory \cite{Bauer:2003pi,Bosch:2004th,Becher:2009th,Becher:2006qw,Becher:2010pd,Bruser:2018rad}. 
Alternatively, as was shown in \cite{Banerjee:2018ozf}, they can be extracted from the coefficient functions of 
DIS with photon and Higgs exchanges \cite{Vermaseren:2005qc,Soar:2009yh}.
Noting that the finite part of $\mathrm \Phi_J^c$ is nothing but the logarithm of Jet function, three loop
contribution to gluon jet function was obtained in \cite{Banerjee:2018ozf}.  

The solution $\mathrm{\Phi}_{J,B}^{c}$ that contains NSV part of the $\mathrm \Phi_J^c$ takes the following form:
\begin{equation}\label{phiB}
    \mathrm{\Phi}_{J,B}^{c}(\hat{a}_s,\mu^2,Q^2,z,\epsilon) = \sum_{i=1}^{\infty} \hat a_s^iS_\epsilon^i \bigg(\frac{Q^2 (1-z)}{\mu^2}\bigg)^{i \frac{\epsilon}{2}}  {1 \over 2} \overline \varphi_c^{(i)}(z,\epsilon) \,.
\end{equation}
We obtain this solution by setting $\overline K^c_J$ to zero and replacing $\overline G_{J}^c$ by 
 $\overline G_J^c - \overline G^c_{J,SV}$ in Eq.(\ref{KGphi}) as they were already taken into account to obtain
SV part of the solution.
The functions $\overline \varphi_c^{(i)}$ contain both UV and IR divergences as poles in $\epsilon$.  The former goes away 
when the coupling constant renormalisation is performed.  
As the entire soft divergences
of real emission processes are contained in  $\Phi^c_{J,A}$,   
the coefficients $\overline \varphi_c^{(i)}(z,\epsilon)$ will have only collinear divergences that will exactly cancel with those of AP kernel.   
The finite part of it can be determined by comparing against $\Delta_c$ order by order in perturbation theory.
We split $\overline \varphi_c^{(i)}$ as a sum of collinear divergent and collinear finite coefficients as
\begin{equation} \label{phisf}
    \overline \varphi_c^{(i)}(z,\epsilon) = \overline \varphi_{s,c}^{(i)}(z,\epsilon) +  \overline \varphi_{f,c}^{(i)}(z,\epsilon) \,.
\end{equation}
From the finiteness of $\Delta_c$ and NSV part of AP kernel, we find that the singular coefficients $\overline \varphi_{s,c}^{(i)}$ is identical to $\overline K_c^{(i)}$ given in Eq.(35) 
of \cite{Ravindran:2006cg} with the following replacement of $A^c$ by $L^c$:
\begin{align}
\label{phisc}
   \overline \varphi_{s,c}^{(i)}(z,\epsilon) = \overline K^{c(i)}_J(\epsilon)\bigg|_{A^{c} \rightarrow L^c(z)} \,,
 \end{align}
where $L^c(a_s(\mu_R^2),z)$ is finite and can be expanded in powers of $a_s(\mu_R^2)$ as
\begin{align}
\label{lc}
   L^c(a_s(\mu_R^2),z)  =  \sum_{i=1}^\infty a_s^i(\mu_R^2) L^c_i(z)
 \end{align}
The coefficients $\overline \varphi_{f,c}^{(i)}(z,\epsilon)$ are determined from NSV terms of $\Delta_c$ 
at every order in perturbation theory.   Although we can determine soft and collinear divergences present in
$\mathrm \Phi_J^c$ at every order in perturbation theory using FF, AP kernel but the finite part requires
the explicit computation of real emission subprocesses around $z =1$.  Note that
$\Delta_c$ are known to third order for several observables in perturbation theory and they allow us 
to extract the finite part of $\mathrm \Phi_J^c$ up to
third order.  In order to determine the finite part, we express the series expansions $\mathrm \Phi_{J,A}^c$ and $\mathrm \Phi_{J,B}^c$ given in eq.(\ref{PhiSV}) and eq.(\ref{phiB}) respectively as
\begin{eqnarray}
\label{phiA}
    \mathrm{\Phi}_{J,A}^{c}(\hat{a}_s,\mu^2,Q^2,z,\epsilon\big) &=
	& \bigg(\frac{1}{2(1-z)}\bigg\{\int_{\mu_F^2}^{Q^2(1-z)} \frac{d\lambda^2}{\lambda^2} A^c(a_s(\lambda^2)) + \overline{G}^c_{J,SV}\big(a_s(Q^2(1-z)),\epsilon\big) \bigg\} \bigg)_+
   \nonumber \\
&&   + \delta(1-z) \sum_{i=1}^\infty \hat a_s^i \bigg(\frac{Q^2}{\mu^2}\bigg)^{i\frac{\epsilon}{2}} 
S_\epsilon^i  {\phi}^{c(i)}_{SV}(\epsilon) 
\nonumber\\ &&
+ {1 \over 2(1-z)_+}\sum_{i=1}^\infty \hat a_s^i  \bigg(\frac{\mu_F^2}{\mu^2}\bigg)^{i\frac{\epsilon}{2}}
S_\epsilon^i  \overline{K}^{c(i)}_J(\epsilon)     .
\end{eqnarray}
where $\overline{G}^c_{J,SV}\big(a_s(Q^2(1-z)),\epsilon\big)$ are related to the threshold exponent $\textbf{B}^I_{DIS}\big(a_s(Q^2(1-z))\big)$ via Eq.(48) of \cite{Ravindran:2006cg}
and 
\begin{eqnarray}
\label{phiBint}
   \mathrm{\Phi}_{J,B}^{c}(\hat{a}_s,\mu^2,Q^2,z,\epsilon\big) &= & \frac{1}{2} \int_{\mu_F^2}^{Q^2(1-z)} \frac{d\lambda^2}{\lambda^2} L^c(a_s(\lambda^2),z) + \overline \varphi_{f,c}\big(a_s(Q^2(1-z)),z,\epsilon\big) |_{\epsilon=0}  \nonumber\\
    &&+ \overline \varphi_{s,c}\big(a_s(\mu_F^2),z,\epsilon\big) \,,
%\sum_{i=1}^{\infty} \hat{a}_s^i \bigg(\frac{\mu_F^2}{\mu^2}\bigg)^{i\frac{\epsilon}{2}}S_{\epsilon}^{i} \overline \varphi_{s,c}^{(i)}\big(z,\epsilon\big)
\end{eqnarray}
where,
\begin{equation}
   \overline \varphi_{a,c}\big(a_s(\lambda^2),z\big) = \sum_{i=1}^\infty
    \hat{a}_s^i \bigg(\frac{\lambda^2}{\mu^2}\bigg)^{i\frac{\epsilon}{2}} S^i_{\epsilon}\frac{1}{2}\overline \varphi_{a,c}^{(i)}\big(z,\epsilon\big) .
\quad \quad \quad a = f,s
\end{equation}
In the expression given in Eq.(\ref{phiBint}) the first line is finite when $\epsilon \rightarrow 0$
whereas second line is divergent in the same limit.  
The RG invariance of $\mathrm{\Phi}_{J,B}^{c}$ implies that $\overline \varphi_{s,c}$ satisfies the renormalisation
group equation:
\begin{eqnarray}
\label{RGphis}
\mu_F^2 {d \over d\mu_F^2}  \overline \varphi_{s,c}(a_s(\mu_F^2),z) = L^c (a_s(\mu_F^2),z).
\end{eqnarray}
The anomalous dimension $L^c$ can be determined by demanding finiteness
of $\Delta_c$ and it turns out that it is half of NSV part of the AP splitting functions (see 
\cite{Ajjath:2020ulr}), that is 
\begin{eqnarray}
L^c(a_s,z) = C^c(a_s) \log(1-z) + D^c (a_s) \,.
\end{eqnarray}  
Note that the SV part of the diagonal splitting function  
in the logarithms of diagonal AP kernel in Eq.(\ref{Psi}) cancels the one from $\overline G^c_{J,SV}$ 
and the remaining divergence coming from NSV part cancels against $\overline \varphi_{s,c}$ making
$\Delta_c$ finite to all orders in $a_s$.  This is guaranteed by the factorisation of collinear divergences
to all orders.

Having understood the structure of the singular part $\overline \varphi_{s,c}$, we now focus on the finite part $\overline \varphi_{f,c}$. The finite part $\overline \varphi_{f,c}$ is parametrised in terms of
$\log^k(1-z)$ as:
\begin{eqnarray}
\label{varphiexp}
\overline \varphi_{f,c}(a_s(Q^2(1-z)),z) = \sum_{i=1}^\infty a_s^i(Q^2(1-z)) \sum_{k=0}^i \frac{1}{2} \overline \varphi_{c,i}^{(k)} \log^k(1-z) \,.
\end{eqnarray}
The coefficients $\overline \varphi_{c,i}^{(k)}$
in Eq.(\ref{varphiexp}) are related to the constants $\mathcal{G}_{L,i}^{c,(j,k)}$s through
\begin{eqnarray}
\overline \varphi_{c,1}^{(k)} &=&  \mathcal{G}_{L,1}^{c,(1,k)}, \quad \quad k=0,1\nonumber\\
\overline \varphi_{c,2}^{(k)} &=&  \frac{1}{2}\mathcal{G}_{L,2}^{c,(1,k)} + \beta_0\mathcal{G}_{L,1}^{c,(2,k)},
k = 0,1,2\nonumber\\
\overline \varphi_{c,3}^{(k)} &=&  \frac{1}{3}\mathcal{G}_{L,3}^{c,(1,k)} + \frac{2}{3}\beta_1\mathcal{G}_{L,1}^{c,(2,k)} + \frac{2}{3}\beta_0\mathcal{G}_{L,2}^{c,(2,k)} + \frac{4}{3}
         \beta_0^2\mathcal{G}_{L,1}^{c,(3,k)}, \quad \quad k=0,1,2,3\nonumber\\
\overline \varphi_{c,4}^{(k)} &=& \frac{1}{4}\mathcal{G}_{L,4}^{c,(1,k)} + \frac{1}{2}\beta_2\mathcal{G}_{L,1}^{c,(2,k)} + \frac{1}{2}\beta_1\mathcal{G}_{L,2}^{c,(2,k)} + \frac{1}{2}
         \beta_0\mathcal{G}_{L,3}^{c,(2,k)}
         + 2\beta_0\beta_1\mathcal{G}_{L,1}^{c,(3,k)}
         + \beta_0^2\mathcal{G}_{L,2}^{c,(3,k)}\nonumber\\
         && +
         2\beta_0^3\mathcal{G}_{L,1}^{c,(4,k)}, \quad \quad k=0,1,2,3,4
\end{eqnarray}
with $\mathcal{G}_{L,1}^{c,(2,3)}$ ,$\mathcal{G}_{L,1}^{c,(2,4)}$,$\mathcal{G}_{L,2}^{c,(2,4)}$,$\mathcal{G}_{L,1}^{c,(3,4)}$ 
not contributing to $\overline \varphi_{c,i}^{(k)}$.  In the above equations,   
$\mathcal{G}_{L,i}^{c,(j,k)}(z)$ are expansion coefficients of
$\mathcal{G}_{L,i}^{c}(z,\epsilon)$ defined by $\overline G_{J,L}^{c} = \overline G_J^c - \overline G^c_{J,SV}$:
\begin{align}
\label{GLc}
\overline G_{J,L}^{c} \Big( \hat{a}_s,\frac{Q^2}{\mu_R^2},\frac{\mu_R^2}{\mu^2},z,\epsilon \Big) &=
\sum_{i=1}^\infty a_s^i\big(Q^2 (1-z)\big) \sum_{j=0}^{\infty} \sum_{k=0}^{i+j-1} \mathcal{G}_{L,i}^{c, (j,k)} \log^k(1-z).
\end{align}
In the dimensionally regularised theory, given the order of perturbation  namely the power of 
$a_s$ and the accuracy of $\epsilon$,  
the loop integrals in the virtual diagrams and 
phase space integrals for the real emission sub-processes demonstrate a systematic logarithmic structure. 
For example, the highest power of
$\log(1-z)$ of the coefficient of $a_s^i \epsilon^j$ in a perturbative expansion, is controlled by both $i$ and $j$.
For the inclusive reactions that we considered, we find that 
the highest power of $\log(1-z)$ is always less than or equal to $i+j-1$.
Hence, the summation over $k$ in the Eq.(\ref{GLc}) runs from 0 to $i+j-1$. 
This translates to the upper limit $i$ in the summation over $k$ in Eq.(\ref{varphiexp}).
At every order $a_s^i$, the coefficients $\mathcal G_{L,i}^{c,(j,k)}$ or their combination namely 
$\overline \varphi_{c,i}^{(k)}$ for various value of
$i$ and $k$ can be extracted from explicit perturbative results of 
$\Delta_c$.
%, $\hat F_c$, $Z_{UV,c}$, $\mathrm{\Phi}_{J,A}^{c}$, and $\Gamma_{c\overline c}$.

So far, we studied how NSV terms can be systematically included in the threshold expansion of inclusive cross section
of DIS.  Same methodology can be applied for the SIA as well to obtain the corresponding all order result.  
Noting that the SIA is time like process, namely the energy scale $Q^2$ is negative of its center of mass energy and
that the collinear factorisation requires time-like splitting functions, we can obtain $\tilde \Psi_J^c$ (see 
Eq.(\ref{Psi}) for SIA by
replacing $Q^2$ in $\hat F_c$ by $-q^2, q^2>0$, $Q^2$ in $\mathrm \Phi_J^c$ by $q^2$ and the splitting functions in $\Gamma_{cc}$
by the time like ones: 
\begin{eqnarray}
\hat F_c(\hat a_s,\mu^2,Q^2,\epsilon) \rightarrow \hat F_c(\hat a_s,\mu^2,-q^2,\epsilon)
\nonumber\\
\mathrm \Phi_J^c(\hat a_s,\mu^2,Q^2,\epsilon) \rightarrow \tilde {\mathrm \Phi}_J^c(\hat a_s,\mu^2,q^2,\epsilon)
\nonumber\\
\ln\Gamma_{cc}(\hat a_s,\mu^2,\mu_F^2,\epsilon) \rightarrow \ln\tilde\Gamma_{cc}(\hat a_s,\mu^2,\mu_F^2,\epsilon)
\end{eqnarray}
where $\tilde \Gamma_{cc}$ is the time like AP kernel.  The solution for jet function for the SIA, 
$\tilde {\mathrm \Phi}_J^c$ is obtained exactly the way we obtained $\mathrm \Phi_J^c$.  
The complete result for the SV part can be found \cite{Blumlein:2006pj} which
we will not repeat here.  For the NSV part of $\tilde{\mathrm \Phi}_{J}^c$, the function $\tilde {\mathrm \Phi}_{J,B}^c$ is found to be the same as Eq.(\ref{phiB}) with the replacements $\overline \varphi_{a,c} \rightarrow \tilde \varphi_{a,c}$ and consequently $\overline \varphi_{a,c} \rightarrow \tilde \varphi_{a,c}$ with $a=s,f$  and $Q^2\rightarrow q^2$.

The coefficient functions $\Delta_c$ for DIS via the exchange of a photon as well as a Higgs boson 
 are available up to third order in \cite{Vermaseren:2005qc,Soar:2009yh} respectively.
One can find the analytical results for FFs, overall renormalisation constants, the functions $\mathrm{\Phi}_{J,A}^{c}$ and 
$\Gamma_{c\overline c}$ up to third order in the literature. 
\textcolor{black}{Following \cite{Moch:2009hr,Amsler:2008zzb}, we define the non singlet DIS structure functions as:
\begin{align}
    {\cal F}_{1} = 2 F_{1,ns} , \qquad  {\cal F}_2 = \frac{1}{x} F_{2,ns}, \qquad {\cal F}_3 = F_{3}^{~ \nu + \bar{\nu}}
\end{align}}
Now using the available results up to third order for $c=q$ (photon-exchange DIS)
% and $c =g $ (Higgs-exchange DIS) \big), 
for the structure functions ${\cal F}_1$ and ${\cal F}_3$, we found 
the functions $\overline \varphi_{q,i}^{(k)}$ as,  
\begin{align}
\label{phiiko}
\overline \varphi_{q,1}^{(0)} &=  10 C_F \,,
%\nonumber\\
\qquad
\overline \varphi_{q,1}^{(1)} =  0\,,
%\nonumber\\
\qquad
\overline \varphi_{q,2}^{(1)} =  10 C_F  C_A   - 22 C_F^2\,,
\nonumber\\
\overline \varphi_{q,2}^{(2)} &=  - 4 C_F^2  \,,
%\nonumber\\
\qquad
\overline \varphi_{q, 3}^{(3)} =  C_F^2 C_A   \bigg(  - {176 \over 27} \bigg) + n_f C_F^2   \bigg( {32 \over 27} \bigg) \,.  %\hspace{6cm}
\end{align}
and for the structure function ${\cal F}_2$,
\begin{small}
\begin{eqnarray}
\label{phiikt}
\overline    \varphi_{q,1}^{(0)} &=&
        14 C_F \,,
%\nonumber\\
\qquad
\overline    \varphi_{q,1}^{(1)} = 0\,,
\nonumber\\
\overline    \varphi_{q,2}^{(0)} &=&
        C_F  C_A    \bigg( {5734 \over 27} - 64  \zeta_3 - {32 \over 3}   \zeta_2 \bigg)
       - C_F^2     \bigg( {25 \over 2} + 44\zeta_2 - 72 \zeta_3 \bigg)
       + n_f  C_F    \bigg(  - {1060 \over 27} + {8 \over 3}  \zeta_2 \bigg)
       \,, \nonumber\\
\overline    \varphi_{q,2}^{(1)} &=&
        C_F  C_A \bigg(-6 + 16\zeta_2 \bigg)  + C_F^2 \bigg(26 - 32\zeta_2 \bigg)  \,,
%\nonumber\\
\qquad
\overline    \varphi_{q,2}^{(2)} =
       - 4 C_F^2  \,,
\nonumber\\ 
\overline \varphi_{q, 3}^{(0)} &=&  C_F C_A^2   \bigg( {3231470 \over 729} + 72 \zeta_5 - {63128 \over 27} 
         \zeta_3 - {46208 \over 81} \zeta_2 - {64 \over 3} \zeta_2 \zeta_3 + {2324 \over 15} 
         \zeta_2^2 \bigg)
\nonumber\\&&
       + C_F^2 C_A   \bigg(  - {83255 \over 54} + {1320} \zeta_5 + {22180 \over 9} 
         \zeta_3 - {25984 \over 27} \zeta_2  + 272 \zeta_2 \zeta_3 -{6464 \over 15} \zeta_2^2 \bigg)
\nonumber\\&&      
       + C_F^3   \bigg( {1319 \over 6} - 2000 \zeta_5 + 444 \zeta_3 -{50 \over 3} \zeta_2 - 224 \zeta_2 \zeta_3
        + {9056 \over 15} 
         \zeta_2^2 \bigg)
%\nonumber\\&&         
       + n_f C_F C_A   \bigg(  - {972940 \over 729}
\nonumber\\&&       
       + {236} \zeta_3 + {21068 \over 81} \zeta_2 - {304 \over 15} \zeta_2^2 \bigg)
%\nonumber\\&&
       + n_f C_F^2   \bigg(  - {133 \over 3} - {880 \over 3} \zeta_3 + {2896 \over 27} 
         \zeta_2 + {256 \over 5} \zeta_2^2 \bigg)
\nonumber\\&&         
       + n_f^2 C_F   \bigg( {68312 \over 729} 
       + {32 \over 27} \zeta_3 - {592 \over 27} 
         \zeta_2 \bigg)
%\nonumber\\&&
         + \bigg[{dabc^{2} \over n}\bigg] fl_{11} \bigg( -128 + 1280 \zeta_5 -704 \zeta_3  - 448 \zeta_2 
\nonumber\\&&         
         + 128 \zeta_2 \zeta_3 + {64 \over 5} \zeta_2^2 \bigg)\,,
\nonumber\\
\overline    \varphi_{q, 3}^{(1)} &=&  C_F C_A^2   \bigg( -{5680 \over 9} + {376 \over 3} \zeta_3 + {1792 \over 3} \zeta_2 - {128 \over 5} \zeta_2^2  \bigg)
       + C_F^2 C_A  \bigg (  +{95612 \over 81} + {1400 \over 3} \zeta_3  - {1004 } \zeta_2 
\nonumber\\&&       
       - {512 \over 5} 
         \zeta_2^2 \bigg)
%\nonumber\\&&
       + C_F^3   \bigg(   {134 \over 3} - 720 \zeta_3 - {160 \over 3} \zeta_2 + {1536 \over 5} \zeta_2^2  \bigg)
       + n_f C_F C_A   \bigg(   {892 \over 9} - {160 \over 3}  \zeta_3 - {260 \over 3} \zeta_2  \bigg)
\nonumber\\&&
       + n_f C_F^2   \bigg( -{16136 \over 81} + {128 \over 3} \zeta_3 + {144} \zeta_2   \bigg) \,,
\nonumber\\
\overline    \varphi_{q, 3}^{(2)} &=&  C_F C_A^2   \bigg( {14 \over 3} - 32 \zeta_3  + 58 \zeta_2 \bigg)
       - C_F^2 C_A   \bigg(   94  - 128 \zeta_3  + {316 \over 3} \zeta_2 \bigg)
       + C_F^3   \bigg( {496 \over 3}  -128 \zeta_3 \bigg)
\nonumber\\&&
       + n_f C_F C_A  \bigg ({2} - {16 \over 3 }\zeta_2  \bigg)
       + n_f C_F^2   \bigg( {4 \over 3}  + {32 \over 3}\zeta_2\bigg) \,,
\nonumber\\
\overline    \varphi_{q, 3}^{(3)} &=&  C_F^2 C_A   \bigg(  - {176 \over 27} \bigg) + n_f C_F^2   \bigg( {32 \over 27} \bigg) \,,
\end{eqnarray}
%Results for eE
\textcolor{black}{Similarly, the non singlet time-like  transverse (${\cal F}_T$) and longitudinal (${\cal F}_L$)  structure functions in SIA
(see \cite{Blumlein:2006pj,Moch:2009hr,Amsler:2008zzb}) are defined as,
\begin{eqnarray}
{\cal F}_T  = F_{T,ns}\,, \quad
{\cal F}_L  = F_{L,ns}\,, \quad
\end{eqnarray}
and $\tilde \varphi_{q,i}^{(k)}$ for  $\cal F_T$ are found to be}
\begin{eqnarray}
{\label{phieECt}}
   \tilde \varphi_{q,1}^{(0)} &=& -8 C_F \,,
%\nonumber\\
\qquad
   \tilde \varphi_{q,1}^{(1)} = 0\,,
%\nonumber\\
\qquad
   \tilde \varphi_{q,2}^{(1)} =
        -10 C_F  C_A   + 22  C_F^2  \,,
\nonumber\\
   \tilde \varphi_{q,2}^{(2)} &=& 4 C_F^2  \,,
%\nonumber\\
  \qquad \ \
   \tilde \varphi_{q, 3}^{(3)} =  C_F^2 C_A   \bigg(   {176 \over 27} \bigg) - n_f C_F^2   \bigg( {32 \over 27} \bigg) \,.
   \hspace{4cm}
\end{eqnarray}
\textcolor{black}{and for  ${\cal F}_L$,  $\tilde \varphi_{q,i}^{(k)}$ are found to be}
\begin{eqnarray}
{\label{phieECL}}
  \tilde  \varphi_{q,1}^{(0)} &=& 2 C_F \,,
%\nonumber\\
\qquad
  \tilde  \varphi_{q,1}^{(1)} = 4 C_F\,,
\nonumber\\
  \tilde  \varphi_{q,2}^{(1)} &=&
        C_F  C_A \bigg({328 \over 9} \bigg)  + C_F^2 \bigg(48 - 16\zeta_2 \bigg) - C_F  n_f \bigg({64 \over 9}\bigg)  \,,
\nonumber\\
  \tilde  \varphi_{q,2}^{(2)} &=&
             {22 \over 3} C_F  C_A  + 8 C_F^2  - C_F  n_f \bigg({4 \over 3}\bigg)\,,
\nonumber\\
  \tilde  \varphi_{q,3}^{(3)} &=&
              C_F  C_A^2 \bigg({484 \over 27} \bigg)  + {88 \over 3} C_A C_F^2 - C_A C_F n_f \bigg({176 \over 27} \bigg) \,             
%\nonumber\\ &&
 - {16  \over 3} C_F^2 n_f + n_f^2 C_F \bigg( {16 \over 27}\bigg).
\end{eqnarray}
\end{small}
%Results for Higgs DIS
and  for $c = g$ (Higgs-exchange DIS) the CFs to the gluon structure function $F_{ \varphi}$ gives the following  
$\overline \varphi_{g,i}^{(k)} $,
\begin{eqnarray} \label{phieg}
    \overline \varphi_{g,1}^{(0)} &=&  {1 \over 3} C_A + {2 \over 3} n_f \,,
%\nonumber\\
\qquad
   \overline \varphi_{g,1}^{(1)} = 0\,,
%\nonumber\\
\qquad
   \overline \varphi_{g,2}^{(1)} =
        -14 C_A^2  + 2 n_f  C_A \,,
\nonumber\\
%\qquad
  \overline \varphi_{g,2}^{(2)} &=&
        -4 C_A^2 \,,
%\nonumber\\
\qquad
   \overline \varphi_{g,3}^{(3)} =
        C_A^3    \bigg(  - {176 \over 27} \bigg)
       + n_f  C_A^2    \bigg( {32 \over 27} \bigg).
       \hspace{3cm}
\end{eqnarray}
Here, the constants $C_A = N_c$ and $C_F = (N_c^2-1)/2 N_c$ are Casimirs of  $SU(N_c)$ gauge group and $n_f$ is number of
active flavours. 
The result for color factor $ \bigg[ {dabc^2 \over N_c} \bigg] fl_{11}$ can be found in \cite{Vermaseren:2005qc}.
For ${\cal F}_1$ and ${\cal F}_3$, we could not obtain all the constants $\overline\varphi_{f,i}^{(k)}$ as
the results for $\Delta_q$ corresponding to them are not available in the literature.  Also for fragmentation
functions, we have given only those that are possible to extract from the available CFs of fragmentation functions. \textcolor{black}{Hence in conclusion, the coefficients $\overline \varphi_{f,q}$ given in Eqs.(\ref{phiiko},\ref{phiikt}) along with the NSV part of the AP splitting functions determine $\mathrm \Phi_{J,B}^q$ up to third order in $a_s$.  Note that $\mathrm \Phi_{J,A}^q$ is already known 
\cite{Ravindran:2006cg,Bauer:2003pi,Bosch:2004th,Becher:2009th,Becher:2006qw,Becher:2010pd,Bruser:2018rad,Banerjee:2018ozf} 
to the same accuracy.  This completes the determination of $\mathrm \Phi_J^c$ to third order in perturbation theory. }   

Having obtained $\mathrm \Phi_J^c$ to third order, we make few observations.  The structure of SV part of $\mathrm \Phi_J^c$, namely $\mathrm \Phi_{J,A}^c$, is
well understood in terms of the cusp anomalous dimension $A^c$ and the function $\overline G_{J,SV}^c$. In particular, one
finds that the entire SV part of $\mathrm \Phi_J^c$ is universal as it is 
independent of the hard interaction.   In the present
case, this means that $\mathrm \Phi_{J,A}^c$ is same for all the structure functions.  However, it depends only on the parton that participates in
the hard scattering.  For photon-DIS, quark and anti-quarks are the ones that interact directly with the photon and hence
the index $c=q,\overline q$ in cusp anomalous dimension and $\overline G_{J,SV}^c$. 
For the Higgs-DIS, both the cusp anomalous dimension as well as $\overline G_{J,SV}^c$ will depend on the gluon and hence they will be
different from those of photon-DIS.  Unlike the SV part, NSV part does not have universal structure even though
part of NSV contains process independent anomalous dimensions $C^c$ and $D^c$ resulting from AP splitting functions.  
From Eqs.(\ref{phiiko},\ref{phiikt}), we find that the explicit results on 
$\overline \varphi_{f,c}$ extracted for different structure functions
do not coincide, implying that they are sensitive to hard scattering of quarks/anti-quarks with the photon.
In \cite{Ajjath:2020ulr}, some of us studied the NSV contributions to production of lepton pairs in Drell-Yan and production of Higgs boson
in bottom quark annihilation and found that the corresponding $\varphi_{f,q}$ and $\varphi_{f,b}$ differ at
third order hinting towards the breakdown of universality for the NSV part.   

%%%%%%%%%%%%%%%%%%%%%%%%%%%%%%%%%%%%%%%%%%%%%%%%%%%%%%%%%%%%%%
\section{All order predictions for $\Delta_c$}

\textcolor{black}{In the earlier section we discussed extensively about each of the building blocks which constitute the master formula given in Eq.\eqref{Psi}. We have also shown that these building blocks satisfy certain differential equations which in turn is controlled by universal anomalous dimensions. Now in this section we aim to discuss the predictability of the solutions to the governing differential equations.} 
% allows us to 
% predict certain logarithms in $\Delta_c$ to all orders in $a_s$. 
For example,
differential equation corresponding to RG can help us to predict 
logarithms of $\mu_R^2$. Similarly AP equations predict logarithms of $\mu_F^2$
and K+G equations of FF and $\mathrm \Phi_J^c$ predict threshold contributions $\delta(1-z)$, ${\cal D}_i$ and
$\log^k(1-z),k=0,1,\cdots$ at higher orders in $a_s$.
Hence, the knowledge of $\mathrm \Phi_J^c$, FFs for the quark/gluon and the AP kernel $\Gamma_{cc}$, all known to third order
can be used to predict certain SV as well as NSV terms in $\Delta_c$ beyond third order.  
Let us expand the CF $\Delta_c$  in powers of $a_s(\mu_R^2)$ as
\begin{eqnarray}
\label{Delexp}
\Delta_c(Q^2,\mu_R^2,\mu_F^2,z) = \sum_{i=0}^\infty a_s^i(\mu_R^2) \Delta_c^{(i)}(Q^2,\mu_R^2,\mu_F^2,z).
\end{eqnarray}
where the coefficient $\Delta_c^{(i)}$ can be determined from Eq.(\ref{MasterF}--\ref{Psi}).
Note that $\Delta_c^{(0)} = \delta(1-z)$.  By definition, $\Delta_c$ contains only SV and NSV terms and hence 
terms of order ${\cal O}((1-z)^\alpha), \alpha > 0$ are dropped.

Using the definition of ${\cal C}$ , we first expand the exponential of $\Psi_J^c$ in Eq.(\ref{MasterF}) in powers of $a_s(\mu_R^2)$ 
and then perform all the convolutions.  This gives, at each order in perturbation theory, a tower of SV terms, 
such as the distributions ${\cal D}_i, i=0,1,\cdots$ and $\delta(1-z)$ and 
of next to SV terms namely the logarithms $\log^i(1-z),i=0,1,\cdots $. 

If $\Psi_J^c$ is known to $a_s$,  
the master formula Eq.(\ref{MasterF}) can predict the leading SV terms 
$({\cal D}_3,{\cal D}_2)$,
$({\cal D}_5,{\cal D}_4),\cdots,({\cal D}_{2i-1},{\cal D}_{2i-2})$ and the leading NSV terms
$\log^3(1-z),\log^5(1-z),\cdots,\log^{2i-1}(1-z)$  at $a_s^2,a_s^3,\cdots,a_s^i$ 
respectively for all $i$.  Note that   
$C^c_1$ is identically zero and hence $\log^{2i}(1-z)$ terms do not contribute irrespective of $i$. 
Similarly the knowledge of $\Psi_J^c$ to order $a_s^2$ can predict
the tower of distributions  $({\cal D}_3,{\cal D}_2), ({\cal D}_5,{\cal D}_4), \cdots,({\cal D}_{2i-3},{\cal D}_{2i-4})$ 
and of $\log^4(1-z),\log^6(1-z),\cdots,\log^{2i-2}(1-z)$ at $a_s^3,a_s^4,\cdots,a_s^i$ respectively for all $i$. 
DIS results for the photon exchange are known to $a_s^3$ and it allows us to confirm our predictions
at second and third orders based on the knowledge of $\mathrm \Phi_J^c$ at $a_s$ and at $a_s^2$ respectively. 
We also confirmed our predictions for SV and NSV terms at third order against
those given in \cite{Vermaseren:2005qc,Soar:2009yh} using the $\mathrm \Phi_J^c$ known to order $a_s$.  This explains the all order predictive nature
of Eq.(\ref{MasterF}).  
The complete knowledge of $\mathrm \Phi_J^c$ up to third order can be used to predict certain SV and NSV terms at fourth order for $\Delta_q$
because the former  
allows us to predict a tower of  $({\cal D}_3,{\cal D}_2),({\cal D}_5,{\cal D}_4) \cdots,({\cal D}_{2i-5},{\cal D}_{2i-6})$
and of $\log^5(1-z),\log^7\cdots,\log^{2i-3}(1-z)$ at $a_s^4,a_s^5,\cdots,a_s^i$ respectively for all $i$. 
%Our predictions for  $\log^7(1-z),\log^6(1-z)$ and $\log^5(1-z)$ terms at fourth order for $\Delta_q$ agree with
%that of  \cite{Vermaseren:2005qc,Soar:2009yh}.  
\begin{table}[h!] \label{tab:Table1}
\begin{center}
\begin{small}
\begin{tabular}{|p{1.2cm}|p{1.5cm}|p{1.5cm}|p{1.5cm}||p{1.5cm}|p{1.5cm}|p{3.15cm}|}
 \hline
 \multicolumn{4}{|c||}{GIVEN} & \multicolumn{3}{c|}{PREDICTIONS}\\
 \hline
 \hline
 \rowcolor{lightgray}
 $\Psi_c^{(1)}$ & $\Psi_c^{(2)}$ &$\Psi_c^{(3)}$&$\Psi_c^{(n)}$&$\Delta_c^{(2)}$&$\Delta_c^{(3)}$& \quad \quad $\Delta_c^{(i)}$\\
 \hline
 ${\cal D}_0,{\cal D}_1,\delta$   &      &&   & ${\cal D}_3,{\cal D}_2$ &${\cal D}_5,{\cal D}_4$ &${\cal D}_{(2i-1)},{\cal D}_{(2i-2)}$\\
% $\delta$ &  & & & &$L_3$ &  ${\cal D}_{(2i-2)}$\\
	$L_{z}^{1},L_{z}^{0}$ &  & &  &$L_{z}^{3}$ & $L_{z}^{5}$&  $L_{z}^{(2i-1)}$\\
 \hline
  &  ${\cal D}_0,{\cal D}_1,\delta$ &   &&&${\cal D}_3,{\cal D}_2$&${\cal D}_{(2i-3)},{\cal D}_{(2i-4)}$\\
%  & $\delta$   & & & & & \\
	& $L_{z}^{2},L_{z}^{1},L_{z}^{0}$ & & & &$L_{z}^{4}$ &$L_{z}^{(2i-2)}$ \\
  \hline
  & & ${\cal D}_0,{\cal D}_1,\delta$& &&&${\cal D}_{(2i-5)},{\cal D}_{(2i-6)}$ \\
%  & &$\delta$   & & $L_3$ &  &\\
	& &$L_{z}^{3},\cdots,L_{z}^{0}$ & & & & $L_{z}^{(2i-3)}$ \\
  \hline
 & & &  ${\cal D}_0,{\cal D}_1,\delta$ &&&${\cal D}_{(2i-(2n-1))},{\cal D}_{(2i-2n)}$  \\
% & &&$\delta$   & &  &$L_3$   \\
	& & & $L_{z}^{n},\cdots,L_{z}^{0}$ & &  &$L_{z}^{(2i-n)}$ \\
  \hline
% & & &  &${\cal D}_0,{\cal D}_1,\delta$ &&  \\
% &&&&$\delta$   & &  \\
%  & & & &$L_n,\cdots,L_0$ &  &  \\
 %\hline
\end{tabular}
\end{small}
\end{center}
	\caption{\label{tab:Table1} Towers of Distributions ($\mathcal{D}_i$) and 
	 NSV logarithms ($\log^i(1-z)$) that can be predicted for $\Delta_c$ using Eq.(\ref{MasterF}). Here $\Psi_c^{(i)}$ and $\Delta_c^{(i)}$ denotes $\Psi_c$ and $\Delta_c$ at order $a_s^i$ respectively. Also the symbol $L_{z}^{i}$ denotes $\log^i(1-z)$.}
\end{table}

In the following, we present our predictions for the NSV terms $L_z$ till seventh order in $a_s$. For the DIS structure function ${\cal F}_1$, we find
\input{DeltaC1.tex}
for the DIS structure function ${\cal F}_2$,
%\begin{eqnarray}
%   \Delta^{(4)}_ {q,2} &=&
%         \textcolor{black}{\Delta^{(4)}_ {q,1}} +  {16 \over 3} C_F^4  \log^6(1-x)
%       +  \bigg(   - {728 \over 9} C_F^3 C_A
%          + 72 C_F^4 
%        + {80 \over 9} n_f C_F^3 + 32 \zeta_2 C_F^3 C_A \hspace{.8cm}
%\nonumber\\&&
%	- 64 \zeta_2 C_F^4 \bigg) \log^5(1-x)\,.
%\end{eqnarray}
\input{DeltaC2.tex}
for the SIA transverse structure function ${\cal F}_T$,
%\begin{eqnarray}
%   \tilde \Delta^{(4)}_ {q,T} &=&
%         \Delta^{(4)}_ {q,1} -  {72 } C_F^4   \log^6(1-x)
%       +  \bigg(  {10316 \over 27} C_F^3 C_A
%          +( 292 - 96 \zeta_2)C_F^4 
% - {2072 \over 27} n_f C_F^3 
%          \bigg) \hspace{.6cm}
%         \nonumber\\&&
%         \log^5(1-x)\,.
%\end{eqnarray}
\input{DeltaCt}
and for the SIA longitudinal structure function ${\cal F}_L$,
%\begin{align}
%   \tilde \Delta^{(4)}_{q,L} =
%           {8 \over 3}C_F^4   \log^6(1-x)
%       -  \bigg( C_F^3 C_A\Big({364 \over 9}  -{16 \zeta_2}\Big) 
%          +4C_F^4 \big(8 \zeta_2-9\big) 
%\nonumber\\&&
%        - {40 \over 9} n_f C_F^3 
%        \nonumber\\&&
%         \bigg) 
 %       \nonumber\\&&
%           \log^5(1-x)\,.
%\end{align}
\input{DeltaCL}
In addition, for the gluon initiated process, we predict
%\begin{eqnarray}
%   \Delta^{(4)}_g &=&
%         \bigg(  - {16 \over 3} C_A^4 \bigg) \log^7(1-x)
%       +  \bigg( -\frac{32}{ 3} C_A^3 n_f + {104} C_A^4 \bigg) 
%\log^6(1-x)
%      +  \bigg(  - {8072 \over 9} C_A^4  
%      \nonumber\\&&
%      + {5132 \over 27} C_A^3 n_f
%           - {176 \over 27}  C_A^2 n_f^2 
%\nonumber\\&&
%        + 176 C_A^4 \zeta_2 \bigg) \log^5(1-x)\,.
%\end{eqnarray} 
\input{Deltag}
Our predictions for  $\log^7(1-z),\log^6(1-z)$ and $\log^5(1-z)$ terms at fourth order for $\Delta_q$s agree with
that of  \cite{Vermaseren:2005qc,Soar:2009yh}.  

In summary, if we know $\Psi_J^c$ up to $n$th order,
we can predict $({\cal D}_{2i-2n+1},{\cal D}_{2i-2n})$ and $\log^{2i-n}(1-z)$ at every order in $a_s^i$ for all $i$, see Table[\ref{tab:Table1}].   
We present the general structure of the NSV partonic CFs  $\Delta_c^{NSV}$ to fourth order in $a_s$ in Appendix[\ref{ap:del}] and  
in the ancillary files.
The fact that the master formula has the predictive nature to all orders in $a_s$ in terms of distributions
and $\log(1-z)$ terms in $\Delta_c$ can be exploited to resum them to all orders.  This will be discussed in the next section.

\section{Resummation in $N$ space}
In the last section, we developed a formalism in $z$ space to study SV and NSV contributions to
DIS and SIA processes.  Our all order result for $\Psi_J^c$ can predict tower of SV logarithms 
for $\Delta_c$ through
the SV distributions ${\cal D}_j$ and NSV logarithms $\log^j(1-z),j=0,1,\cdots$ at every order in $a_s$.
This is possible because the knowledge of SV and NSV terms for $\Delta_c$ 
up to a given order $m$ in $a_s$, say $a_s^i, i=0,\cdots,m$, contains valuable information 
through $\Psi_J^c$, for LL, NLL etc at every order in $a_s^j$ with $j>m$, 
i.e., for terms at $a_s^j, j=m,\cdots ,\infty$.  The reason for this is due to the fact that the 
constituents of $\Psi_J^c$, namely
the FF, $Z_{UV,c}$, $\mathrm \Phi_J^c$ and AP kernels, satisfy differential equations and  their perturbative 
solutions have all order predictions for certain logarithms, such as logarithms of $Q^2$, 
$\mu_R^2$, $\mu_F^2$ and also logarithms of the form $\log^j(1-z)/(1-z)^k,j=0,1,\cdots,\infty, k=0,1$.  
The structure of these logarithms is controlled by UV and IR anomalous dimensions.
Presence of such logarithms is a unique feature of any perturbative expansions and is considered
advantageous to evaluate whether the perturbative series is reliable or not.  
They can also become large at every order posing problem for the perturbative series.   
For example, the distributions ${\cal D}_j$ and the NSV terms  $\log^j(1-z)$ 
in threshold region can become large at every order in $a_s$.
In practice, there are  situations when the order of perturbation increases,
the contributions from distributions ${\cal D}_j$ also increase such that the product
$a_s^j {\cal D}_{j-1}$ is of order one at every order $j=1,\cdots,\infty$. 
This means that we need to take into account order one terms to all orders in perturbation theory
to make any sensible prediction.

Since we are dealing with distributions in $\Psi_J^c$ and convolutions in $\Delta_c$, it is convenient to
work in Mellin space $N$.  The distributions in $N$ space are well defined. Also the convolutions
in $z$ space become normal products in $N$ space.  Hence, it is easy to study the order
one terms in $N$ space.
The threshold limit 
namely $z\rightarrow 1$  where the SV distributions and NSV logarithms $\log^j(1-z)$  dominate,  
corresponds
to $N \rightarrow \infty$ in Mellin space.  We do not strictly take $N\rightarrow \infty$ as we are interested
in NSV terms.  In $N$ space, in the large $N$ limit, the Mellin moment of $z$ space SV 
gives $\log^i(N)$ terms as well as $1/N^j \log^i(N)$ terms.  However, the $z$ space NSV terms
will always give $1/N^j \log^i(N)$ terms.  While performing Mellin moments, we drop $1/N^j \log^i(N)$ terms
with $j>1$ for all $i$ at every order in $a_s$.   This way, we have in $N$ space, SV contributions
which contain only $\log^i(N)$ terms whereas NSV contains only $1/N \log^i(N)$ term.  
The order one terms that we found in $z$ space in the threshold limit  
will show up in the Mellin space through $(a_s \beta_0 \log(N))^k$ with $k>0$. 
Reorganisation of the perturbative series taking into account these order one terms consistently 
can be achieved through the procedure called resummation. 
In the rest of the section,  we show how these order one terms in $\Psi_J^c$ 
can be summed to all orders in perturbation theory. 
The resummed results in Mellin space taking into account only SV terms are available 
in the literature for variety of processes, 
see \cite{Catani:1996yz,Moch:2005ba,Bonvini:2012an,Bonvini:2014joa,Bonvini:2014tea,Bonvini:2016frm, Bonvini:2016fgf,H:2019dcl,Bonvini:2010ny,Bonvini:2012sh, H.:2020ecd,Catani:2014uta,Cacciari:2001cw}.  In the following,
we derive the N space resummed result for the NSV logarithms.  Unlike the resummed SV contributions, 
the NSV terms organise themselves in double series expansion in both $a_s$ as well as in $\log(N)$. In addition,
we find that the resummed expression for NSV part is $1/N$ suppressed compared to SV. 
     
It is convenient to use the integral representations of $\mathrm{\Phi}_{J,A}^{c}$ and 
$\mathrm{\Phi}_{J,B}^{c}$ given in Eq.(\ref{phiA}) and Eq.(\ref{phiBint}) respectively to perform
the Mellin moment in the large $N$ limit to obtain SV $\log^i(N)$ terms as well as NSV $1/N \log^i(N)$ terms. 
Substituting Eq.(\ref{phiA}) and Eq.(\ref{phiBint}) in $\Delta_c$, the Mellin moment of Eq.(\ref{MasterF}) takes the following form,  
\begin{eqnarray}
\label{DeltaN}
\Delta_{c,N}(Q^2,\mu_R^2,\mu_F^2) = C_0^c(Q^2,\mu_R^2,\mu_F^2) \exp\left(
\Psi^c_{J,N} (Q^2,\mu_F^2) 
\right)\,,
\end{eqnarray}
$\Psi^c_{J,N}$ in the above equation  is twice the Mellin moment of $\Psi_{J,\cal D}^c$,
\begin{eqnarray}
\Psi^c_{J,N}(Q^2,\mu_F^2) = 2 \int_0^1 dz z^{N-1}\Psi_{J,\cal D}^c (Q^2,\mu_F^2,z)\,. 
\end{eqnarray}
where
\begin{eqnarray}
\label{phicint}
\Psi^c_{J,\cal D} (Q^2,\mu_F^2,z) &=& {1 \over 2}
\int_{\mu_F^2}^{Q^2 (1-z)} {d \lambda^2 \over \lambda^2} 
{ P}^{\prime}_{cc} (a_s(\lambda^2),z)  + {\cal Q}^c(a_s(Q^2 (1-z)),z)\,,
\end{eqnarray}
%with
%\begin{eqnarray}
%\label{calPcc}
%	{\cal P}_{cc}\big(a_s,z\big) &=&   A^c(a_s) {\cal D}_0(z) + L^c(a_s,z)   \,,
%\end{eqnarray}
 and
 \begin{eqnarray}
\label{calQc}
{\cal Q}^c (a_s(Q^2(1-z)),z) &=&  \left({1 \over 2(1-z)} \overline G^c_{J,SV}(a_s(Q^2 (1-z)))\right)_+ 
%\nonumber\\&& 
+ \overline \varphi_{f,c}(a_s(Q^2(1-z)),z)\,.
\end{eqnarray}
The $N$ independent coefficient $C_0^c$ results from the finite parts of  FF, $\Gamma_{cc}$ and the
coefficient of  $\delta(1-z)$ of  $\mathrm \Phi^c_J$.  We expand $C_0^c$ in powers
of $a_s$ as:  
\begin{eqnarray}
\label{C0expand}
C_0^c(Q^2,\mu_R^2,\mu_F^2) =  \sum_{i=0}^\infty a_s^i(\mu_R^2) C_{0i}^c(Q^2,\mu_R^2,\mu_F^2)\,,
\end{eqnarray}
where the coefficients $C^c_{0i}$ are presented in the ancillary files. 
Also,the results of $C^c_{0}$ for the photon-exchange DIS can be found in \cite{Moch:2005ba}. 
Both $C^c_0$ and $\Psi^c_{J,N}$ in Eq.(\ref{DeltaN}) depend on process dependent 
quantities as well as the universal anomalous dimensions, $A^c,B^c,C^c,D^c,f^c$ and  
set of SV coefficients $\overline G^c_{J,SV}$ and NSV coefficients $\overline \varphi_{f,c}$. 

{\color{black}Before we proceed to the details of computation, 
%the computation of the Mellin moments of the exponent in Eq.(\ref{DeltaN}),we want to
we make} few remarks on the $N$ space result that we obtained.  There have been several studies on
the all order structure of NSV terms, see 
\cite{Laenen:2008ux,Laenen:2010kp,Laenen:2010uz,Bonocore:2014wua,Bonocore:2015esa,Beneke:2019oqx,Beneke:2019mua,Bonocore:2016awd,Beneke:2018gvs,Bahjat-Abbas:2019fqa} 
in order to better understand the underlying IR physics. We find that our result given in Eq.(\ref{DeltaN}) is 
similar to the one which was conjectured in \cite{Laenen:2008ux}.  However, we differ
from Eq.(37) in \cite{Laenen:2008ux}, in the upper limit of the integral, the presence of extra term 
$\overline \varphi_{f,c}$ and the explicit dependence on the variable $z$. 
While these differences do not disturb the SV predictions, they will give NSV terms different from those obtained using Eq.(37) of \cite{Laenen:2008ux}. 

Our next task is to perform the Mellin moment of SV and NSV terms,
%keeping terms up to order
%${\cal O}(1/N)$ and drop all those that are of order ${\cal O}(1/N^i)$ for $i>1$.
where we keep all terms till ${\cal O}(1/N)$ and drop the rest which have higher power in 1/N. 
We encounter two types of integrals, namely integrals over distributions ${\cal D}_i,i=0,1,\cdots$ and
those over regular terms $\log^i(1-z),i=0,1,\cdots$.  Care is needed to deal with the integrals of
distributions.  The section 2.2 of \cite{Laenen:2008ux} contain results that are suitable to 
obtain Mellin moments of distributions as well as regular terms in the large $N$ limit.   
Following, \cite{Laenen:2008ux}, we replace $\int dz ({z^{N-1}-1})/(1-z)$ and $\int dz z^{N-1}$ by 
$\int \theta(1-z-1/N)/(1-z)$ and apply the operators $\Gamma_A(N \frac{d}{dN})$ and $\Gamma_B(N \frac{d}{dN})$ on 
them respectively.  The $\Gamma_A\left(N{d \over dN}\right)$ and $\Gamma_B\left(N{d \over dN}\right)$ 
are given by
\begin{eqnarray}
\label{GAGB}
   \Gamma_A\big(x\big) =  \sum\limits_{k=0} - \gamma_{k}^Ax^k, \quad \quad
   \Gamma_B\big(x\big) = \sum\limits_{k=1}\gamma_{k}^Bx^{k},
\end{eqnarray}
%\begin{eqnarray}
%\label{GAGB}
%\Gamma_K &=& \sum_{k=0}^\infty  \gamma_i^K
%\left(N{d \over dN}\right)^k
%\,,\quad \quad K = A,B
%\end{eqnarray}
\begin{eqnarray}
\gamma_k^A = {\Gamma_k(N) \over k!} (-1)^{k-1} \,, \quad \quad 
\gamma_{k+1}^B = {\Gamma^{(k)}(1) \over k!} (-1)^{k}. 
\end{eqnarray}
and they are listed in the Appendix[C] of \cite{Ajjath:2020ulr}. 
Applying these operators, we find 
\begin{eqnarray}
\label{PsiNres}
   \Psi_{J,N}^c &=& 
	\int^{Q^2}_{Q^2/N} \frac{d\lambda^2}{\lambda^2} \bigg\{
	     \bigg(\ln\frac{\lambda^2 N}{Q^2} + \gamma_{1}^A \bigg) A^c(a_s(\lambda^2)) - \overline{G}^c_{J,SV}\big(a_s(\lambda^2)\big) -\frac{1}{N} \xi^{c}(a_s(\lambda^2),N)
\nonumber\\&&
	- \lambda^2 {d \over d \lambda^2} \mathcal{F}^c (a_s(\lambda^2),N) \bigg\} 
+ \mathcal{F}^c(a_s(Q^2),N)
\nonumber\\&&
     + \int_{\mu_F^2}^{Q^2} \frac{d\lambda^2}{\lambda^2} \bigg\{ \bigg(-\gamma_{1}^A - \log(N)\bigg)  A^c(a_s(\lambda^2)) + \frac{1}{N} \xi^c(a_s(\lambda^2),N) \bigg\} \,,
\end{eqnarray}
where 
\begin{eqnarray}
\label{F}
\mathcal{F}^c (a_s,N) = \mathcal{F}^c_A (a_s) + \frac{1}{N} \mathcal{F}^c_B (a_s,N).
\end{eqnarray}
with $\mathcal{F}^c_A(a_s)$ and $\mathcal{F}^c_B (a_s,N)$ are found to be 
\begin{eqnarray}
	\mathcal{F}_A^c(a_s) & =& - \gamma_{1}^A  \overline{G}^c_{J,SV}\big(a_s\big) 
	+ \sum_{i=0}^\infty \gamma_{{i+2}}^A \Big(- \beta (a_s)\frac{\partial}{\partial a_s}\Big)^i 
	\bigg\{   A^c(a_s) 
%\nonumber\\&&
	+ \beta(a_s) {\partial \over \partial a_s}  
	\overline{G}^c_{J,SV}\big(a_s\big) \bigg \},
\nonumber\\
\end{eqnarray}
and 
\begin{eqnarray}
  \mathcal{F}^c_B\big(a_s,N\big) &=& 
2\gamma_{1}^B \overline \varphi_{f,c}(a_s,N) - 2 \gamma_{2}^B\bigg( \lambda^2 \frac{d}{d\lambda^2} \overline \varphi_{f,c}(a_s,N) 
+ \frac{1}{2}\tilde\xi^c(a_s,N)\bigg) \nonumber\\
   &&  + 2 \bigg( \gamma_{3}^B + \tilde{\gamma}^{B}\bigg) \bigg(\lambda^2\frac{d}{d \lambda^2}
\bigg\{
\lambda^2\frac{d}{d \lambda^2}  \overline \varphi_{f,c}(a_s,N) 
+ \frac{1}{2 }\tilde \xi^c(a_s,N) \bigg\}
%\nonumber\\
%&&
+ \frac{1}{2} C^c(a_s)\bigg).
\nonumber\\
\end{eqnarray}
% \begin{eqnarray}
%   \mathcal{F}^c_B\big(a_s,N\big) &=& 
% 2\gamma_{1}^B \overline \varphi_{f,c}(a_s,N) - 2 \gamma_{2}^B\bigg( \beta(a_s) \frac{d}{da_s} \overline \varphi_{f,c}(a_s,N) 
% + \frac{1}{2}\tilde\xi^c(a_s,N)\bigg) \nonumber\\
%   &&  + 2 \bigg( \gamma_{3}^B + \tilde{\gamma}^{B}\bigg) \bigg(\beta (a_s)\frac{d}{d a_s}
% \bigg\{
% \beta(a_s)\frac{d}{da_s}  \overline \varphi_{f,c}(a_s,N) 
% + \frac{1}{2 }\tilde \xi^c(a_s,N) \bigg\}
% %\nonumber\\
% %&&
% + \frac{1}{2} C^c(a_s)\bigg),
% \nonumber\\
% \end{eqnarray}
In the above equation, $ \tilde \gamma^B = \sum^\infty\limits_{i=4} \gamma_{i}^B (N \frac{d}{d N})^{i-3} $. For brevity we denote $a_s(\lambda^2)$  as $a_s$ in all the above equations. Also,
\begin{eqnarray}
\tilde\xi^c(a_s,N) &=& 
D^c(a_s) -C^c(a_s) \log(N)
%\nonumber\\
\,, \quad
\xi^c(a_s,N) =  \gamma_{1}^B
\tilde\xi^c(a_s,N) 
%\bigg( D^c(a_s) - C^c(a_s) \log(N) \bigg) 
- \gamma_{2}^BC^c(a_s).
\end{eqnarray}
What remains to be done now is the integration 
over $\lambda^2$ in Eq.(\ref{PsiNres}) or equivalently over $a_s$.  
Care is needed to keep the order one term, namely terms of the form $\omega^j,j=1,\cdots,\infty$, with 
$\omega=a_s(\mu_R^2) \beta_0 \log(N)$, at every order in $a_s(\mu_R^2)$.      
If we use RG equation for $a_s$ and replace integration over $\lambda^2$ by $a_s$, we obtain 
the  result that is expected to be a function of 
$a_s(Q^2/N)$ as well as $a_s(Q^2)$.  Both these $a_s$s can be   
expanded around $a_s(\mu_R^2)$ using RG of the strong coupling constant keeping order one $\omega$s intact.
Alternatively, we can achieve this by performing the integrations over $\lambda$ by using $a_s(\lambda^2)$ given by 
\begin{eqnarray}
\label{resumas}
a_s(\lambda^2)&=&\bigg({a_s(\mu_R^2) \over l}\bigg) \bigg[1-{a_s(\mu_R^2)\over l} {\beta_1 \over \beta_0} \log(l)
   + \bigg({a_s(\mu_R^2) \over l}\bigg)^2 \bigg({\beta_1^2\over \beta_0^2} (\log^2(l)-\log(l)
\nonumber\\
&&+l-1)-{\beta_2\over \beta_0} (l-1) \bigg)
   + \bigg({a_s(\mu_R^2)\over l}\bigg)^3 \bigg({\beta_1^3\over \beta_0^3} \Big(2 (1-l) \log(l) + {5\over 2} \log^2(l)
\nonumber \\
&&- \log^3(l) -{1\over 2} + l - {1\over 2} l^2\Big)
             +{\beta_3 \over 2 \beta_0} (1-l^2) + {\beta_1 \beta_2 \over \beta_0^2} \Big(2 l \log(l)
\nonumber\\
&&- 3 \log(l) - l (1-l)\Big)\bigg)\bigg ].
\end{eqnarray}
where $l = 1 - \beta_0 a_s(\mu_R^2) \log(\mu_R^2/\lambda^2)$ and $\beta_i$ are the coefficients of QCD beta function 
$\beta(a_s) =-\sum\limits_{i=0}^\infty a_s^{i+2} \beta_i$, known to five loops, see 
\cite{Chetyrkin:2017bjc,Luthe:2017ttg,Herzog:2017ohr,Baikov:2016tgj}.
%\cite{vanRitbergen:1997va,Czakon:2004bu}. 
The latter approach is easier to retain order one $\omega$s at every order in $a_s(\mu_R^2)$ and we followed this
in our paper.  
Since $\overline \varphi_{f,c}$ depends both on $a_s$ and $\log(N)$, we make a double series expansion as follows:
\begin{eqnarray}
\label{varphiexpN}
	\overline \varphi_{f,c}(a_s,N) &=& \sum \limits_{i=1}^{\infty} \sum \limits_{k=0}^{i} a_s^i \frac{1}{2} \overline \varphi_{c,i}^{(k)} (-\log(N)) ^k\,.
\end{eqnarray}
using the expansion for $\overline \varphi_{f,c}$ and 
expanding other quantities in the exponent in powers of $a_s$,
and using Eq.(\ref{resumas}) for $a_s(\lambda^2)$ in Eq.(\ref{PsiNres}).
We obtain 
\begin{align}
\label{PsiN}
\Psi_{J,N}^c =
&\log(g_0^c(a_s(\mu_R^2))) + \tilde g_1^c(\omega)\log(N) 
+ \sum_{i=0}^\infty a_s^i(\mu_R^2) \tilde g_{i+2}^c(\omega)
%\nonumber\\
%&
\nonumber\\&
 +\frac{1}{N} 
 %\bigg(
%\bar g_1^c(\omega) \log(N) + 
%\sum_{i=0}^\infty a_s^i(\mu_R^2) \bar g_{i+2}^c(\omega) +    
\sum_{i=0}^{\infty} a_s^i(\mu_R^2) h^c_{i}(\omega,N)\,,
%\bigg) \,,
\end{align}
where $\tilde g_i^c$ and $h_{i}^c(\omega,N)$ are defined as,
\begin{align}
\label{hg}	
 &\tilde g_i^c(\omega) = g_i^c(\omega) + \frac{1}{N}\overline{g}_i^c(\omega) \,,
 \nonumber \\
	h^c_0(\omega,N) = h^c_{00}(\omega) + h^c_{01}(\omega) &\log(N),
	\quad h^c_i(\omega,N) = \sum_{k=0}^{i} h^c_{ik}(\omega)~ \log^k(N) .
\end{align}
Here we have dropped terms order ${\cal O}(1/N^i),i>1$.
Also, the function $\log(g_0^c(a_s))$ is expanded in powers of $a_s$ as,
\begin{align}
\log(g_0^c(a_s(\mu_R^2))) = \sum_{i=1}^\infty a_s^i(\mu_R^2) g_{0,i}^c \,.
\end{align}
In Eq.(\ref{PsiN}), the coefficients  $g^c_0$ and $g^c_i$,$i =1,2, \cdots$ correspond to SV part, 
whereas $\overline g_i^c$ and $h_i^c$,$i=1,2,\cdots$ correspond to NSV part. 
The coefficients $g^c_i(\omega)$ are identical to those in \cite{Moch:2005ba} obtained from 
the resummed formula for SV terms. Note that $g^c_i(\omega)$ becomes zero in the 
limit $\omega \rightarrow 1$. 
The coefficient $g^c_0(a_s)$ (see \cite{Moch:2005ba})
is given in the ancillary files. 
The $N$ independent coefficients $C^c_{0}$ and $g^c_{0}$ can be combined as 
\begin{align}
    \tilde g^c_0(Q^2,\mu_R^2,\mu_F^2) = C^c_0(Q^2,\mu_R^2,\mu_F^2) g^c_0(a_s(\mu_R^2)),
\end{align}
and is expanded in terms of $a_s(\mu_R^2)$ as,
\begin{align}
    \tilde g^c_0(a_s(\mu_R^2)) = \sum_{i=0}^\infty a_s^i(\mu_R^2) \tilde g^c_{0,i}\quad \,.
\end{align}
Here, $\tilde g^c_{0,i}$ are found to be identical to $g^{DIS}_{0k}$ given in \cite{Moch:2005ba}. 
While both $z$ space as well as $N$ space results contain same information, the summation of
order one $\omega$ terms to all orders in perturbation theory can be conveniently performed only 
in $N$ space.  We find that the functions $g_i^c,\bar g_i^c$ and $h_i^c$ sum up $\omega$ to all orders in perturbation theory.

In Eq.(\ref{PsiN}), $\bar g ^c_1$ is found to be identically zero and the remaining terms   
$\bar g^c_i, i= 2,3,\cdots$ are functions of the cusp anomalous dimension $A^c$ and 
the function $\overline G_{J,SV}^c$ of Eq.(\ref{calQc}). $\bar g^c_i(\omega)$ contain no explicit $\log(N)$ terms. 
The functions $h_i^c$ in Eq.(\ref{PsiN}) result from the Mellin moment of
$\mathrm \Phi^c_{J,B}$ and hence depend  on the anomalous dimensions $C^c,D^c$ and the function 
$\overline \varphi_{f,c}$.
We find that the coefficient $h^c_{01}$ is proportional to $C^c_1$ which is identically zero, and hence 
 at order $a_s^0$, there is no $(1/N) \log(N)$ term.  

% We do not present explicit results of $g^c_i,\bar g^c_i$ and $h_i^c$ here as they have similar
% form as those contributing to the resummation of inclusive DY or Higgs productions, \cite{Ajjath:2020ulr}, instead
% we provide the translations that are required to obtain them from the later ones.

%\textcolor{red}{We find that there exist certain transformation rules following which one can obtain the resummation coefficients $g^c_i,\bar g^c_i$ and $h_i^c$ from the 

 We also observe from the explicit calculations, that there exist certain transformation rule which relates to the resummation coefficients $\tilde g^c_i$ and $h_i^c$ of DIS with the corresponding coefficients of DY/Higgs production presented in \cite{Ajjath:2020ulr}. These rules are found to be,
\begin{align}
    \tilde g_1^{DY}(\omega) &= 2 \tilde g_1^{DIS}(2\omega), \quad \tilde g_{i+1}^{DY}(\omega) = \tilde g_{i+1}^{DIS}(2\omega)\Big|_{\substack{\big\{B_i\rightarrow0,\gamma_i^A\rightarrow 2^i \gamma_i^A\big\}}},\quad i>0\,,
    \nonumber \\
    h_{ik}^{DY}(\omega,N) &= 2^k h_{ik}^{DIS}(2\omega,N)\Big|_{\substack{\big\{D_i\rightarrow 2D_i,\gamma_i^B\rightarrow 2^{(i-1)} \gamma_i^B,\overline \varphi_i^{(k)}\rightarrow2^{(1-k)}\overline \varphi_i^{(k)}\big\}}}.
\end{align}
%where, 
%\begin{align}
%    \tilde g_i^{c,I} = g_i^{c,I}(\omega) + \frac{1}{N} \bar g_i^{c,I}(\omega),\quad \forall \quad i>0 .
%\end{align}
%}
% The coefficients $g^c_i(\omega)$ and $\bar g^c_i(\omega)$ can be obtained from the corresponding ones given in  \cite{Ajjath:2020ulr}    
% by setting $B_i^c=0$ and replacing $\gamma_A^i$ by $2^i \gamma_A^i$ for $i=1,2,3$ and $g_1^c$ by twice of $g_1^c$.
% For the coefficients $h_{ik}^c$, we need to replace $\gamma_B^i$ by $2^{i-1} \gamma_B^i$ for $=1,\cdots,4https://www.overleaf.com/project/5f0d709490a8c40001b30482#equation.3.13$, 
% $\varphi_{c,i}^{(k)}$ by $2^{1-j}\overline \varphi_{c,i}^{(k)}$ and multiply $1/2^k$ with each $h_{ik}^c$. 
\textcolor{black}{The reason behind the existence of such transformation rule follows from the fact that $\omega$ is related to $\log N$ in DIS whereas for DY/Higgs it is related to $\log N^2$. This dependency resurfaces at every stages of the calculation. The scaling of $\gamma_i^A$($\gamma_i^B$) reflects the same dependency as they are coefficients of the derivative of $\log N$ in the expansion of $\Gamma_A$ ($\Gamma_B$) defined in Eq.(\ref{GAGB}). Also $\overline \varphi_i^{(k)}$, being coefficients of $\log N$ (see Eq.(\ref{varphiexpN})), gets scaled accordingly. Hence the above relations between DIS and DY/Higgs manifest the aforementioned dependency.  }
The results of the coefficients $g_i^c(\omega)$ are given in \cite{Das:2019btv} and $\bar{g}_i^c(\omega)$ and $h_{ik}(\omega,N)$ for DIS up to four loops are given in Appendix[\ref{ap:hij}, \ref{ap:gbar}]. For completeness we also provide them in the ancillary files.

Let us now study the all order structure of the exponent in the $N$ space. First of all, it is
clear that working in $N$ space in the large $N$ limit (keeping order ${\cal O}(1/N)$ terms) 
allows us to cast the entire exponent in a compact form through the functions $g_i^c,\bar g_i^c$ and $h_i^c$,
each of which is a function of $\omega$.
These functions carry all order information of SV and NSV logarithms.  
This is not surprising because $\mathrm{\Phi}_J^c$ does contain the same information in $z$ space,
however, with no compact looking form.  
Note that in $z$ space we have the inherent scale $Q^2 (1-z)$ of the process appearing at every order
with $\hat a_s$ through $\hat a_s^i (Q^2(1-z))^{i \epsilon/2}$ term. 
This results in $a_s(Q^2 (1-z))$ through UV renormalisation which demonstrates the all order prediction
of logarithm through distributions and $\log(1-z)$  when expanded around $a_s(\mu_R^2)$.  In $N$ space, these logarithms can be
systematically summed up at every order $a_s(\mu_R^2)$ through the functions $g_i^c,\bar g_i^c$ and $h_i^c$ given in the exponent.
We find that unlike SV part, NSV part is a double series expansion in $a_s(\mu_R^2)$ and $\log(N)$.  The explicit 
$\log(N)$s in the NSV functions $h_i^c$ comes from explicit $\log(1-z)$ terms in the expansion of $\overline \varphi_{f,c}$ and
these logarithms do not demonstrate any all order structure and hence can not be summed like the ones appear in the argument of 
$a_s$.  
Note that the index $i$ in the functions $g_i^c$ etc that appears in the exponent   
determine the accuracy of SV and NSV logarithms that are summed up to all orders.         
Hence, expanding the $\mathrm \Phi_{J,A}^c$ or its Mellin moment around $a_s(\mu_R^2)$  
allows us to make definite predictions for SV and NSV logarithms with given accuracy to all orders
in $a_s(\mu_R^2)$.  For example if we have $\mathrm \Phi_{J,A}^c$ up to order $a_s$, 
we can predict terms of the form $a_s^i {\cal D}_{2i-1}(z)$ in $\mathrm{\Phi}_{J,A}^c$ for all $i>1$.
The functions $\tilde g_{0,0}$ and $g_1$ in the exponent can predict leading $a_s^i \log^{2i} (N)$ terms
for all $i>1$.  If we include $\tilde g_{0,1}$ and $g_2$ terms, then we can predict
next to leading $a_s^i \log^{2i-1} (N)$ terms for all $i>2$.
In general, the resummed
result with terms {\color{black} $\tilde g^c_{0,0},\cdot\cdot \cdot \tilde g^c_{0,n-1}$ and $g^n_1,\cdot \cdot \cdot, g^c_{n}$ can predict
$a_s^i \log^{2i-n+1}(N)$.}
%\vspace{5mm}
The inclusion of sub leading terms through $\bar g_i^c$ and $h_i^c$
we get additional $(1/N) \log^j(N)$ terms in $N$ space or $\log^j(1-z)$ terms in $z$ space.
In perturbative QCD, $C^c_1=0$, where $c=q,\overline q, g$.
Like the SV part of the resummed exponent, the $1/N$ suppressed terms also 
organises themselves by keeping
$\omega=a_s(\mu_R^2) \beta_0 \log(N)$ terms as order one at every order in $a_s(\mu_R^2)$. 
In addition, we find that at a given order $a_s^i(\mu_R^2)$, the $1/N$ coefficient is a polynomial in $\log(N)$ 
with the order $i$. 
Again, we find that using {\color{black} $\tilde g^c_{0,0},g^c_1,g^c_2$  and  $\bar g^c_1,\bar g^c_2, h^c_{0},h^c_1$}, one
can predict $(a_s^i/N) \log^{2i-1}(N)$ terms for all $i>1$.
Similarly, along with the previous ones, {\color{black} $\tilde g^c_{0,1}$,$g^c_3$  and $\bar g^c_3,
 h^c_2$ }, one
can predict $(a_s^i /N) \log^{2i-2} (N)$ for all $i>2$.
This way, the resummed result
with $\bar g^c_2 , \cdots,\bar g^c_{n+1}$ and $h^c_1 , \cdot \cdot \cdot ,h^c_{n}$  along with
${\color{black} \tilde g_{0,0},\cdot\cdot \cdot, \tilde g_{0,n-1}}$ and $g_1,\cdot \cdot \cdot, g_{n+1}$  can
predict $(a_s^i/N) \log^{2i - n}(N)$ for all $i>n$ in Mellin space $N$. 
To illustrate the above discussion, we compare the three loop predictions obtained from $\tilde g_{0,i-1}$,$g_{i+1}^c$, $\bar g_{i+1}^c$ and $h_i^c$ for $i<3$ against the exact three loop results in Table[\ref{tab:Table2}]. It can be seen that, given the previous order results, all the higher logarithmic coefficients can be exactly predicted. However,
lower order $1/N \log^k(N),k<4$ can not be predicted from our all order result as they require $\overline \varphi_{c,3}^{(k)}$ for $k<4$ which can only be determined
from the third order result for CFs. Interestingly, even without the knowledge of these terms, our predictions for $1/N \log^3(N)$ terms agree with the
exact result for several color factors, see Table [\ref{tab:Table2}]. Note that the limitations in the predictions for higher orders
from the previous order for NSV terms are in close resemblance with those of
 higher order predictions for SV terms in CF,  given lower order SV exponents.
 In summary, our all order result,
both in $z$ space and $N$ space demonstrates all order structure as well as  predictions that have identical features for
both SV and NSV terms.
The only difference between SV and NSV terms in the exponent is the way they depend on the process.  That is, we find that
NSV exponents are process dependent unlike SV ones. Table [2] of \cite{Ajjath:2020ulr} summarizes these observations at any given order.
\begin{table}[H] \label{tab:Table2}
\begin{center}
\begin{small}
%\newcolumntype{P}[1]{>{\centering\arraybackslash}p{#1}}
{\renewcommand{\arraystretch}{1.7}
\begin{tabular}{|p{1.5cm}||P{1.2cm}|P{1.2cm}||P{1.2cm}|P{1.2cm}||P{2.3cm}|P{2.8cm}|}
 
 \hline
 \rowcolor{lightgray}
    & \multicolumn{2}{c||}{$\frac{\log^5(N)}{N}$}  &  \multicolumn{2}{c||}{$\frac{\log^4(N)}{N}$}  
    & \multicolumn{2}{c|}{$\frac{\log^3(N)}{N}$}  \\ 
  \hline
  \hline
%  $C_F^3$   &\vtop{\hbox{\strut   $\frac{256}{27}$}\hbox{\strut}}   &   $\frac{256}{27}$
    $C_F^3$   & $\frac{256}{27}$   &   $\frac{256}{27}$
  &   
   $ \frac{6499}{ 40}$   &   ${6499 \over 40}$   &   
   ${167031 \over 500} + {3584  \over 27}\zeta_2$   &    ${167031 \over 500} + {3584  \over 27}\zeta_2$ \\ 
 % \hline  
   %  $C_F^2 n_f$&\vtop{\hbox{\strut $0$}\hbox{\strut}}  & 0 & $-{1600\over 81}$ & 
          $C_F^2 n_f$&$0$  & 0 & $-{1600\over 81}$ &
     $-\frac{1600}{81}$ & 
   $-{431451 \over 1000}$ & $-{105229 \over 250} + \chi_1$  \\ 
 %  \hline
   % $C_A C_F^2$   &\vtop{\hbox{\strut$0$} \hbox{\strut }}   &   $0$   &  
    $C_A C_F^2$   &$0$   &   $0$   & 
    $\frac{1760}{27}$   &     $\frac{1760}{27}$   &   $\frac{38617}{25}-{256\zeta_2}$   &   $\frac{150991}{100}-256\zeta_2 + \chi_2$  \\ 
   
 %  $C_F n_f^2$   &\vtop{\hbox{\strut$0$}\hbox{\strut }}   &   $0$   &   $0$   &   $0$   &
     $C_F n_f^2$   &$0$  &   $0$   &   $0$   &   $0$   &
   $\frac{800}{81}$   &   $\frac{800}{81}$   \\ 
 
  % $C_A C_F n_f$   &\vtop{\hbox{\strut$0$}\hbox{\strut }}   &   $0$   &  
   $C_A C_F n_f$   &$0$   &   $0$   &  
   $0$   &   
   $0$   &   
   $-\frac{1760}{27}$   &   $-\frac{1760}{27}$   \\ 
%    $C_A^2 C_F $   &\vtop{\hbox{\strut$0$}\hbox{\strut }}   &   $0$   &   $0$   &   $0$  
    $C_A^2 C_F $   &$0$   &   $0$   &   $0$   &   $0$  
    &   $\frac{968}{9}$   &   $\frac{968}{9}$  \\     
  \hline
\end{tabular}}
\end{small}
\end{center}
	\caption{\label{tab:Table2}: Comparison of 3-loop resummed predictions against exact results for the DIS structure function $F_2.$ For each color structure, the left column stands for the exact results and the right column stands for the resummed predictions. The constant $\chi_1$ and $\chi_2$ depends on $\overline \varphi_{q,3}^{(3)}$.}
\end{table}

\section{Physical evolution kernel}
So far, we studied the structure of CFs in the threshold limit taking into account both the dominant SV and  sub dominant 
NSV terms.  Recall that we work in a dimensionally regularised quantum field theory where all the divergences
are regularised in $4+\epsilon$ dimensions and use modified minimal subtraction scheme to 
perform both UV renormalisation as well as mass factorisation.   Hence, the CFs and PDF or PFF depend
on this scheme, hence they are unphysical.  However, the physical observables comprising of them are blind to this.  
This means that if we make scheme transformations on CFs and on PDFs or PFF simultaneously, the physical
observables are invariant.  For example, following \cite{Blumlein:2000wh},
we consider an observable ${\cal O}(Q^2) ={\cal C}_{\cal O} (Q^2,\mu_F^2) {\cal F}(\mu_F^2)$,
where ${\cal O}$ represent any of the structure functions that appear in the DIS or in the cross section
for SIA.  The functions ${\cal C}_{\cal O}$ and ${\cal F}$ are the corresponding CFs and PDFs or PFFs respectively.  They are 
independently scheme dependent quantities.
That is, if we make scheme transformations, namely  ${\cal C}_{\cal O} \rightarrow  Z {\cal C}_{\cal O}$ and
${\cal F} \rightarrow {\cal F}/Z$, then ${\cal O}$ remains invariant.  This fact allows us to construct 
perturbative quantities out of CFs and PDF/PFF such that they are invariant under scheme transformation.   
One such quantity is ${\cal K}_{{\cal O}}$, called physical evolution kernel (PEK) \cite{Curci:1980uw,Floratos:1980hm}
and is defined through 
\begin{eqnarray}
\label{calO}
Q^2 {d \over dQ^2}{\cal O}(Q^2) = {\cal K}_{{\cal O}}(Q^2) {\cal O}(Q^2)\,,
\end{eqnarray}
where, ${\cal K}_{{\cal O}}$ is obtained using the RG equation for ${\cal C}_{\cal O}$ and is given by 
\begin{eqnarray}
\label{calKO}
{\cal K}_{{\cal O}}(Q^2) = \gamma_{{\cal O}}(Q^2) + Q^2 {d C_{\cal O}(Q^2)\over dQ^2}C^{-1}_{\cal O}(Q^2)\,.
\end{eqnarray}
The anomalous dimension $\gamma_{\cal O}$ satisfies 
$\mu_F^2 d/d\mu_F^2 (\log({\cal F}_{\cal O}(\mu_F^2))=\gamma_{\cal O}(\mu_F^2)$.
Being scheme independent, the kernel can be used to understand the perturbative
structure of physical quantities.  In \cite{Blumlein:2000wh}, the crossing relation namely
the Drell-Levy-Yan relation \cite{Drell:1969jm} between CFs of DIS and of 
SIA were studied in a scheme invariant way using PEK.   
In \cite{Grunberg:2009yi,Grunberg:2009vs}, next to NSV to DIS was studied using PEK. 

A striking observation was made by Moch and Vogt
in \cite{Moch:2009hr} (and \cite{deFlorian:2014vta,Das:2020adl}),
by studying PEKs of observables in DIS, semi-inclusive $e^+ e^-$ annihilation and DY.
They showed that the PEKs demonstrate the enhancement of a single-logarithms at large $z$ to all order in $1-z$.
Making use of this observation and extending it to all orders in $a_s$, the structure of corresponding
leading $\log(1-z)$ terms in the kernel can be constrained, which allowed them to predict
certain next to SV logarithms at  higher orders in $a_s$.   

Now that we have an all order results for SV+NSV both in $z$ and $N$ spaces, 
we can easily predict the structure of leading logarithms in the physical evolution kernel.
It is convenient to use the result in $N$ space for this purpose.  The PEK for DIS is given by 
\begin{eqnarray}
{\cal K}^c(a_s(\mu_R^2),N) &=&  Q^2 {d \over d Q^2} \log \Delta_{c,N}(Q^2) .
\end{eqnarray}
which is invariant under scheme transformation.   
The kernel ${\cal K}^c(a_s(\mu_R^2),N)$ can be computed  order by order in perturbation theory using $\Delta_{c,N}$,
\begin{eqnarray}
	{\cal K}^c(a_s(\mu_R^2),N) = \sum_{i=1}^\infty a_s^i(\mu_R^2)  {\cal K}^c_{i-1}(N).
\end{eqnarray}
As in \cite{Moch:2009hr}, the leading $(1/N) \log^i(N)$ terms at every order
defined by $ \overline {\cal K}^c$:
\begin{eqnarray}
\overline {\cal K}_i^c = {\cal K}_i^c\bigg|_{(1/N) \log^i(N) }
\end{eqnarray}
can be obtained.   
Using Eq.(\ref{DeltaN}), we find that these terms can be obtained directly from the NSV part of $\Psi_{J,N}^c$ alone 
and are given by 
\begin{align}
\label{PEK}
\overline{\cal K}^c_i &= \frac{1}{2} A_1^c \beta_0^i + D_1^c \beta_0^i - i \beta_0^{(i-1)}  C_2^c 
+ i \sum_{j=1}^i (-1)^{j+1} \binom{i-1}{j-1}  \beta_0^{(i-j+1)}~ \overline{\varphi}_{c,j}^{(j)} \,.
\end{align}

The corresponding PEKs for SIA can be obtained from the above DIS PEKs 
by replacing $D^c_1,C^c_2$ and $\overline \varphi_{c,i}^{(i)}$ by
the respective time-like counter parts.   
%the following PEKs:
%\begin{eqnarray}
%\overline {\cal K}^c_0 &=& - 2 C_F \,,
%\nonumber\\
%\overline {\cal K}^c_1 &=&  -2 \beta_0 C_F \pm 16 C_F^2\,,
%\nonumber\\
%\overline {\cal K}^c_2&=& -2 \beta_0^2 C_F \pm 24 \beta_0 C_F^2\,,
%\nonumber\\
%\overline {\cal K}_3&=& -2 \beta_0^3 C_F \pm 24 \beta_0^2 C_F^2 - {176 \over 9} \beta_0 C_A C_F^2 + { 32 \over 9} \beta_0 C_F^2 n_f 
%\,,\nonumber\\
%\overline {\cal K}^c_4&=& -2 \beta_0^4 C_F \pm 16 \beta_0^3 C_F^2 - {704 \over 9} \beta_0^2 C_A C_F^2  + {128 \over 9} \beta_0^2 C_F^2 n_f - 4 \beta_0~ \overline \varphi_{c,4}^{(4)}.
%\nonumber
%\end{eqnarray}
Interestingly at every order in $a^i_s$ the leading $(1/N) \log^i(N)$ terms are controlled
by $D^c_1$ and $C^c_2$ from $L^c$ and by $\overline \varphi_{c,i}^{(i)}$.  This is our prediction for 
the leading $1/N$ behavior of the physical evolution kernel, $\overline {\cal K}^c$, at every order in perturbation theory  and 
is consistent with the observation made by \cite{Moch:2009hr} using the known perturbative results.
%We find a complete numerical agreement with \cite{Moch:2009hr} up  to three loops for all the structure functions
%in photon-exchange DIS and fragmentation functions in SIA
%when we substitute the known values of $C^c,D^c$ of space and time-like splitting functions 
%and DIS NSV constants  $\overline \varphi_{c,i}^{(i)}$ and SIA ones $\tilde \varphi_{c,i}^{(i)}$ from 
%Eqs.  (\ref{phiiko},\ref{phiikt},\ref{phieECt},\ref{phieECL}) respectively. 

% \overline {\cal K}^g_0 &=& - 2 C_A \,,
% \nonumber\\
% \overline {\cal K}^g_1 &=&  -2 \beta_0 C_A - 16 C_A^2\,,
% \nonumber\\
% \overline {\cal K}^g_2&=& -2 \beta_0^2 C_A - 24 \beta_0 C_A^2\,,
% \nonumber\\
% \overline {\cal K}^g_3&=& -2 \beta_0^3 C_A - 24 \beta_0^2 C_A^2 - {176 \over 9} \beta_0 C_A^3 + { 32 \over 9} \beta_0 C_A^2 n_f
% \,,\nonumber\\
% \overline {\cal K}^g_4&=& -2 \beta_0^4 C_A - 16 \beta_0^3 C_A^2 - {704 \over 9} \beta_0^2 C_A^3  + {128 \over 9} \beta_0^2 C_A^2 n_f - 4 \beta_0~ \overline \varphi_{g,4}^{(4)}.
% %\nonumber
% \end{eqnarray}
For photon-exchange DIS and fragmentation functions in SIA we find a complete numerical agreement with \cite{Moch:2009hr} up  to three loops for all the structure functions
%in photon-exchange DIS and fragmentation functions in SIA
upon substituting the known values of $C^c,D^c$ of space and time-like splitting functions
and DIS NSV constants  $\overline \varphi_{c,i}^{(i)}$ and SIA ones $\tilde \varphi_{c,i}^{(i)}$ from
Eqs.  (\ref{phiiko},\ref{phiikt},\ref{phieECt},\ref{phieECL}) respectively. 
%For completeness we present the result for both DIS and SIA in the following:
In the following we present the PEKs for the fragmentation functions in SIA, photon-exchange DIS and Higgs-exchange DIS:
\begin{align}
\label{PEK2}
\overline {\cal K}^c_0 =& - 2 C_i \,,
\nonumber\\
\overline {\cal K}^c_1 =&  -2 \beta_0 C_i \pm 16 C_i^2\,,
\nonumber\\
\overline {\cal K}^c_2=& -2 \beta_0^2 C_i \pm 24 \beta_0 C_i^2\,,
\nonumber\\
\overline {\cal K}^c_3=& -2 \beta_0^3 C_i \pm {88 \over 3} \beta_0^2 C_i^2\,,
\nonumber \\
\overline {\cal K}^c_4=& -2 \beta_0^4 C_i \pm {112 \over 3} \beta_0^3 C_i^2  - 4 \beta_0~ \overline \varphi_{c,4}^{(4)}\,,
\nonumber \\
\overline {\cal K}^c_5=&   -2 \beta_0^5 C_i \pm {160 \over 3} \beta_0^4 C_i^2  - 20 \beta_0^2~ \overline \varphi_{c,4}^{(4)} + 5 \beta_0~ \overline \varphi_{c,5}^{(5)}  \,,
\nonumber \\
\overline {\cal K}^c_6=&   -2 \beta_0^6 C_i \pm {248 \over 3} \beta_0^5 C_i^2  - 60 \beta_0^3~ \overline \varphi_{c,4}^{(4)} + 30 \beta_0^2~ \overline \varphi_{c,5}^{(5)} - 6 \beta_0~ \overline \varphi_{c,6}^{(6)}. 
\end{align}
Here $C_i $ is $\{C_A,C_F\}$ for $c=\{g,q\}$ respectively. And the plus sign corresponds to SIA and the minus sign is for photon and Higgs-exchange DIS.
%\textcolor{black}{Here plus sign corresponds to SIA and the minus sign is for photon-exchange DIS. 
This owe to the fact that the anomalous dimension $C_2$ is equal and opposite in sign for space-like and time-like splitting kernels. Moreover from Eq.(\ref{phiikt},\ref{phieECt},\ref{phieg}) we  can also see that $\overline \varphi_{q,k}^{(k)}$ from two loop onwards are equal and opposite in sign giving rise to the above differences in sign between SIA and DIS.

The agreement of our predictions for the leading term in the kernel with those obtained using explicit results
for CFs is not surprising because the K+G equation that the $\mathrm \Phi^c_J$ satisfies and the
functions $\overline G_{J,L}^c$s are similar to physical evolution equation and the PEK respectively.   
The highest power of $\log(N)$ in the $1/N$ coefficient of $\overline {\cal K}^c$ is due to the 
upper limit on the summation in Eq.(\ref{varphiexpN}). \textcolor{black}{We make another interesting comparison between Eq.(\ref{PEK}) and Eq.(\ref{PEK2}). The general structure for PEKs in terms of the anomalous dimensions and $\overline \varphi_{c,k}^{(k)}$ given in Eq.(\ref{PEK}) shows that  $\overline {\cal K}^c_j$ at any given order is proportional to $\beta_0^i$, $i=j,\cdots,1$. On the other hand for Eq.(\ref{PEK2}) we have coefficients only proportional to $\beta_0^i$ and $\beta_0^{i-1}$ for $\overline {\cal K}^c_i$. Hence the substitution of the explicit values of $A_1,C_2,D_1$ and $\overline \varphi_{c,k}^{(k)}$ in Eq.(\ref{PEK}) conspires in such a way so as to keep only $\beta_0^i$ and $\beta_0^{i-1}$ at every order, which calls for an explanation.}
However, this exercise provides a consistency check on our framework.
Detailed study on the structure of sub leading contributions
to the PEKs, namely, the coefficients of $1/N \log^{i-1}(N),1/N \log^{i-2}(N)$ etc at every order $a_s^i$ can unravel the $\log(N)$ 
structure of $\overline \varphi_{f,c}$. 
\textcolor{black}{It is to be noted that the results for $\overline {\cal K}_4^c$, $\overline {\cal K}_5^c$ and $\overline {\cal K}_6^c$ are incomplete 
%in terms of $\overline \varphi_{c,k}^{(k)}$. %At present, we could not give $\overline %\varphi_{c,k}^{(k)}$ for $k = 4,5$ and $6$ 
due to the unavailability of the full explicit NSV results for the CFs at $a_s^4$, $a_s^5$ and $a_s^6$ orders.}

While the fixed order predictions for CFs of DIS and SIA from our formalism  agree with those \cite{Moch:2009hr}, our formalism provides a result in $z$ that sums up both SV and NSV logarithms to all orders in $a_s$ in terms of certain 
universal anomalous dimensions and process dependent constants.
This is possible because we could get an all order solution
for the K+G equation that results from factorisation and renormalisation group invariance.  In addition, our $z$ space
result allows to perform resummation of order one terms in the
exponent to improve the precision of theoretical predictions from
the NSV terms.

\section{Conclusions}
Perturbative structure of the observables in QFT is very rich and the higher order predictions for them
provide enormous opportunities to unravel the details of the underlying dynamics.  In particular,
inclusive observables such as structure functions in DIS and fragmentation functions in SIA can be used
to understand factorisation properties in QCD perturbation theory.  This requires the knowledge of
infrared structure of on-shell amplitudes at higher orders.  IR structure of perturbative results
show universality through process independent anomalous dimensions, splitting functions, soft distributions and jet 
functions etc.  The universality of these quantities allows us to use the results from one process to make predictions
for the other ones.  For example, quantities like IR anomalous dimensions, soft functions parametrise resummation exponents
in a process independent way allowing us to use them for a class of processes to perform threshold resummation.   
Often, threshold corrections are important in certain kinematical regions and they contribute  
significantly at every order spoiling the reliability of perturbation theory.   Resummation of such
contributions to all orders provides a possible solution.  The most dominant one, namely SV contributions are resummed, 
conveniently in Mellin space.
They show up as order one contributions through $\omega=a_s \beta_0 \log(N)$ at every order in perturbation theory.
Next to SV terms are often not small at every order demanding careful investigation of their structure.
Such an investigation would help us to systematically include them to all orders.  

In this paper, we begin by studying the structure of $\log^k(1-z),k=0,\cdots,\infty$ NSV terms at every order $a_s^i$.  
We used collinear factorisation of inclusive cross sections, their RG invariance and
K+G equations for the form factors and for the real emission contributions that contribute.   
We obtain an all order result in $z$ space that describes the IR structure of NSV terms in terms of
IR anomalous dimensions $C^c$ and $D^c$ and the functions $\overline \varphi_{f,c}$.  
Given the result for CF to specific order in $a_s$, 
our $z$ space result can predict SV and NSV terms of CF with certain accuracy to all orders in $a_s$.  This feature
is the result of universal IR structure that these contributions demonstrate. 

In addition, we find that the finite part of soft distribution and next to soft distribution functions 
have integral representations that can be used to study them in Mellin $N$ space. 
In Mellin $N$ space taking into account large $\log(N)$ and $1/N \log(N)$ contributions, one can
easily resum order one terms namely $\omega$ to all orders in $a_s$.  We also find that
$\overline \varphi_{f,c}$s depend on the hard scattering process breaking the universality that
the SV part of the inclusive cross sections enjoy.  Like in $z$ space, if we know CFs to certain order in $a_s$, 
our results can predict certain $1/N$ suppressed $\log(N)$ contributions
to all orders in $a_s$.  While this paper addresses only quark initiated reactions (gluons in Higgs-DIS) both in 
DIS and SIA, the  NSV terms resulting from other scattering reactions can not be ignored and the perturbative
structure and possible resummation of them to all orders are need of the hour to obtain any consistent analysis. 
The resummed result taking into account $1/N$ corrections will be useful to study the phenomenological 
importance of NSV contributions to inclusive observables.  
\section{Acknowledgements}
We thank Moch for useful discussion, his constant help throughout this project and also for providing third order 
results for the inclusive reactions. 
\appendix
\input{appendix}

\bibliographystyle{JHEP}
\bibliography{nsvdis}
\end{document}

%% file: DeltaC1.tex
\begin{align}
\begin{autobreak}
   \Delta^{NSV}_{q,1} = 
        a_s  \Delta^{NSV (1)}_{q,1} 
        + a_s^2  \Delta^{NSV (2)}_{q,1} 
        +  a_s^3\Delta^{NSV (3)}_{q,1} 
        + {\color{blue} \pmb{a_s^4}} \bigg[
        \bigg\{  - {16 \over 3} C_F^4 \bigg\} L_z^7 +  \bigg\{ {308 \over 9} C_F^3 C_A  
        + {232 \over 3} C_F^4 -  {56 \over 9} n_f C_F^3 \bigg\}  L_z^6
        -  \bigg\{   {1936 \over 27} C_F^2 C_A^2 
        + \bigg( {16384 \over 27}
        - 48 \zeta_2 \bigg) C_F^3 C_A 
        + \big( 188  - 128 \zeta_2 \big) C_F^4  - {704 \over 27} n_f C_F^2 C_A 
        - {2920 \over 27} n_f C_F^3 + {64 \over 27} n_f^2 C_F^2 \bigg\}  L_z^5 +  \mathcal{O}\Big( {\color{black}{ L^4_{z}}} \Big) \bigg]
        
        +  {\color{blue} \pmb{a_s^5}} \bigg[  \bigg\{ - \frac{8}{3} C_F^5
        \bigg\}  L_z^9 +           \bigg\{ - \frac{16}{3} C_F^4 n_f
        + \frac{88}{3} C_F^4 C_A
        + \frac{148}{3} C_F^5  \bigg\}  L_z^8 +
        \bigg\{  - \frac{320}{81} C_F^3 n_f^2
        + \frac{3520}{81} C_F^3 C_A n_f
        - \frac{9680}{81} C_F^3 C_A^2
        + \frac{3056}{27} C_F^4 n_f
        + \bigg( - \frac{17216}{27}
        + \frac{128}{3} \zeta_2  \bigg) C_F^4 C_A 
        + \bigg( - \frac{532}{3}
        + \frac{320}{3} \zeta_2 \bigg) C_F^5 \bigg\}  L_z^7 +    \mathcal{O}\Big( {\color{black}{ L^6_{z}}} \Big) \bigg]

         +  {\color{blue} \pmb{a_s^6}}  \bigg[ \bigg\{- \frac{16}{15} C_F^6
         \bigg\}  L_z^{11} + \bigg\{
          - \frac{88}{27} C_F^5 n_f
          + \frac{484}{27} C_F^5 C_A
          + 24 C_F^6
          \bigg\}   L_z^{10}
          + \bigg\{
          - \frac{320}{81} C_F^4 n_f^2
          + \frac{3520}{81} C_F^4 C_A n_f
          - \frac{9680}{81} C_F^4 C_A^2
          + \frac{6632}{81} C_F^5 n_f
          + \bigg( - \frac{37376}{81} 
          
          + \frac{80}{3} \zeta_2 \bigg) C_F^5 C_A
          + \bigg(- \frac{344}{3} +  64 \zeta_2 \bigg) C_F^6 \bigg\}  L_z^9 
          +  \mathcal{O}\Big( {\color{black}{ L^8_{z}}} \Big) \bigg]
    + {\color{blue} \pmb{a_s^7}}  \bigg[ \bigg\{- \frac{16}{45} C_F^7 \bigg\}  L_z^{13}
         + \bigg\{ - \frac{208}{135} C_F^6 n_f
          + \frac{1144}{135} C_F^6 C_A + \frac{424}{45} C_F^7 \bigg\}   L_z^{12}  
          +   \bigg\{ - \frac{224}{81} C_F^5 n_f^2 + \frac{2464}{81} C_F^5 C_A n_f
          - \frac{6776}{81} C_F^5 C_A^2
          + \frac{18128}{405} C_F^6 n_f
           + \bigg( - \frac{20416}{81}
           + \frac{64}{5} \zeta_2 \bigg) C_F^6 C_A
          + \bigg( - \frac{844}{15}
          + \frac{448}{15} \zeta_2 \bigg)  C_F^7  \bigg\}  L_z^{11}  +  \mathcal{O}\Big( {\color{black}{ L^{10}_{z}}} \Big) \bigg ]\,,
\end{autobreak}
\end{align}

%% file: DeltaC2.tex
\begin{align}
\begin{autobreak}

\end{autobreak}\nonumber\\
\begin{autobreak}
   \Delta^{NSV}_{q,2} =
   a_s  \Delta^{NSV (1)}_{q,2} + a_s^2  \Delta^{NSV (2)}_{q,2} +  a_s^3
   \Delta^{NSV (3)}_{q,2} +  {\color{blue} \pmb{a_s^4}} \bigg[ \Delta^{NSV (4)}_{q,1} +
       \bigg\{ {16 \over 3} C_F^4 \bigg\}  L_z^6
       - \bigg\{    {728 \over 9} C_F^3 C_A
          
        - {80 \over 9} n_f C_F^3 - 32 \zeta_2 C_F^3 C_A + \big( - 72 
	+ 64 \zeta_2 \big) C_F^4 \bigg\}  L_z^5 +  \mathcal{O}\Big( {\color{black}{ L^4_{z}}} \Big)\bigg] 
   
+ {\color{blue} \pmb{a_s^5}} \bigg[ \Delta^{NSV (5)}_{q,1} +
        \bigg\{
           \frac{8}{3} C_F^5
          \bigg\} L_z^8

       +   \bigg\{
           \bigg( - \frac{544}{9}
          + \frac{64}{3} \zeta_2 \bigg) C_F^4 C_A 
          + \frac{64}{9} C_F^4 n_f
          + \bigg ( 48 
          - \frac{128}{3}\zeta_2 \bigg) C_F^5 
          \bigg\}  L_z^7   +  \mathcal{O}\Big( {\color{black}{ L^6_{z}}} \Big) \bigg] + 
{\color{blue} \pmb{a_s^6}} \bigg[  \Delta^{NSV (6)}_{q,1} +
          \bigg\{
           \frac{16}{15} C_F^6
          \bigg\}  L_z^{10}

       +    \bigg\{
         \bigg(  - \frac{904}{27} C_F^5 C_A
          + \frac{32}{3} \zeta_2 \bigg) C_F^5 C_A 
          + \frac{112}{27} C_F^5 n_f
          + \bigg( 24 
          - \frac{64}{3} \zeta_2 \bigg) C_F^6 
          \bigg\} L_z^9   +  \mathcal{O}\Big( {\color{black}{ L^8_{z}}} \Big) \bigg] +
          
          {\color{blue} \pmb{a_s^7}} \bigg[ \Delta^{NSV (7)}_{q,1} 

       +  \bigg\{ \frac{16}{45} C_F^7  \bigg\}  L_z^{12} 

       +    \bigg\{ \bigg( - \frac{1984}{135} 
          + \frac{64}{15} \zeta_2 \bigg) C_F^6 C_A 
          + \frac{256}{135} C_F^6 n_f
          + \bigg( \frac{48}{5} C_F^7 
          - \frac{128}{15} \zeta_2 \bigg) C_F^7 
          \bigg\}  L_z^{11}  +  \mathcal{O}\Big( {\color{black}{ L^{10}_{z}}} \Big) \bigg]\,,
\end{autobreak}
\end{align}

%% file: DeltaCt.tex
\begin{align}
\begin{autobreak}
\end{autobreak}\nonumber\\
\begin{autobreak}
   \tilde \Delta^{NSV}_{q,T} =
   a_s  \tilde \Delta^{NSV (1)}_{q,T} 
   + a_s^2  \tilde \Delta^{NSV (2)}_{q,T} 
   +  a_s^3 \tilde \Delta^{NSV (3)}_{q,T} 
   +  {\color{blue} \pmb{a_s^4}} \bigg[ \Delta^{NSV (4)}_{q,1}  -  {72 } C_F^4   L_z^6
   +  \bigg\{  {10316 \over 27} C_F^3 C_A +\big( 292 - 96 \zeta_2\big)C_F^4 - {2072 \over 27}  n_f C_F^3 \bigg\}
         L_z^5   +  \mathcal{O}\Big( {\color{black}{ L^4_{z}}} \Big) \bigg] 

  + {\color{blue} \pmb{a_s^5}} \bigg[ \Delta^{NSV (5)}_{q,1}   
  -  44 C_F^5  L_z^8
       +  \bigg\{ \frac{11072}{27} C_F^4 C_A
          - \frac{2144}{27} C_F^4 n_f  + \Bigg( \frac{728}{3} 
          - 64 \zeta_2 \bigg)C_F^5 \bigg\} L_z^7 +  \mathcal{O}\Big( {\color{black}{ L^6_{z}}} \Big) \bigg] 

   + {\color{blue} \pmb{a_s^6}} \bigg[ \Delta^{NSV (6)}_{q,1}  -   \bigg\{ \frac{104}{5} C_F^6  \bigg\} L_z^{10} +   \bigg\{   \frac{24484}{81} C_F^5 C_A  - \frac{4648}{81} C_F^5 n_f
          + \bigg( \frac{436}{3} 
          - 32  \zeta_2 \bigg) C_F^6  \bigg\} L_z^9 +  \mathcal{O}\Big( {\color{black}{ L^8_{z}}} \Big) \bigg]

  + {\color{blue} \pmb{a_s^7}} \bigg[ \Delta^{NSV (7)}_{q,1}

          - 8 C_F^7 L_z^{12}

       +  \bigg\{\frac{67888}{405} C_F^6 C_A
          - \frac{12736}{405} C_F^6 n_f
          + \bigg( \frac{1016}{15} 
          - \frac{64}{5} \zeta_2 \bigg) C_F^7  \bigg\} L_z^{11}  +  \mathcal{O}\Big( {\color{black}{ L^{10}_{z}}} \Big) \bigg]\,,

\end{autobreak}
\end{align}

%% file: DeltaCL.tex
\begin{align}
\begin{autobreak}
\end{autobreak}\nonumber\\
\begin{autobreak}
   \tilde \Delta^{NSV}_{q,L} =
       a_s  \tilde \Delta^{NSV (1)}_{q,L} 
       + a_s^2  \tilde \Delta^{NSV (2)}_{q,L} 
       +  a_s^3 \tilde \Delta^{NSV (3)}_{q,L} 
       +  {\color{blue} \pmb{a_s^4}} \bigg[ 
       \bigg( {8 \over 3}C_F^4  \bigg) L_z^6
       -  \bigg\{  \bigg({364 \over 9}  -{16 \zeta_2}\bigg)C_F^3 C_A 
        + \big(32 \zeta_2-36\big) C_F^4 
        - {40 \over 9} n_f C_F^3 
        \bigg\}  L_z^5  +  \mathcal{O}\Big( {\color{black}{ L^4_{z}}} \Big) \bigg]

        +  {\color{blue} \pmb{a_s^5}} \bigg[ \bigg\{ \frac{4}{3} C_F^5 \bigg\} L_z^8  
        +  \bigg\{ \frac{32}{9} C_F^4 n_f 
        +\bigg(  24 - \frac{64}{3} \zeta_2 \bigg) C_F^5   + \bigg(  - \frac{272}{9}+ \frac{32}{3}
        \zeta_2 \bigg) C_A C_F^4 \bigg\}  L_z^7  
        +  \mathcal{O}\Big( {\color{black}{ L^6_{z}}} \Big)\bigg]  + {\color{blue} \pmb{a_s^6}} \bigg[ \bigg\{ \frac{8}{15} C_F^6 \bigg\} L_z^{10} 
        + \bigg\{ \frac{56}{27} C_F^5 n_f   + \bigg( 12- \frac{32}{3} \zeta_2 \bigg) C_F^6  + \bigg( - \frac{452}{27}   + \frac{16}{3} \zeta_2  \bigg) C_A C_F^5 \bigg\} L_z^9 +  \mathcal{O}\Big( {\color{black}{ L^8_{z}}} \Big) \bigg] 
        + {\color{blue} \pmb{a_s^7}} \bigg[ \bigg\{ \frac{8}{45} C_F^7 \bigg\} L_z^{12} +  \bigg\{ \frac{128}{135} C_F^6 n_f  + \bigg( \frac{24}{5} -  \frac{64}{15} \zeta_2 \bigg) C_F^7 - \bigg(  \frac{992}{135} 
        -  \frac{32}{15} \zeta_2 \bigg)  C_A C_F^6 \bigg\}  L_z^{11}  +  \mathcal{O}\Big( {\color{black}{ L^{10}_{z}}} \Big) \bigg].

\end{autobreak}
\end{align}

%% file: Deltag.tex
\begin{align}
\begin{autobreak}
     \Delta^{NSV}_g = 
       a_s \Delta^{NSV (1)}_g  
       +  a_s^2 \Delta^{NSV (2)}_g 
       +  a_s^3 \Delta^{NSV (3)}_g  
       +  {\color{blue} \pmb{a_s^4}} \bigg[  \bigg\{  - {16 \over 3} C_A^4 \bigg\} L_z^7   -  \bigg\{ \frac{32}{ 3} C_A^3 n_f 
       - {104} C_A^4 \bigg\} L_z^6 +  \bigg\{ {5132 \over 27} C_A^3 n_f   -{176 \over 27}  C_A^2 n_f^2 + \bigg(  - {8072 \over 9}  + 176\zeta_2 \bigg) C_A^4  \bigg\} L_z^5  
       +  \mathcal{O}\Big( {\color{black}{ L^4_{z}}} \Big) \bigg] + {\color{blue} \pmb{a_s^5}} \bigg[ \bigg\{  - \frac{8}{3} C_A^5 \bigg\} L_z^9 
   
       + \bigg\{  - \frac{76}{9} C_A^4 n_f + \frac{682}{9} C_A^5 \bigg\} L_z^8   - \bigg\{ \frac{800}{81} C_A^3 n_f^2 
       - \frac{18088}{81} C_A^4 n_f  - \bigg( \frac{78488}{81} - \frac{448}{3} \zeta_2 \bigg) C_A^5 \bigg\} L_z^7  +  \mathcal{O}\Big( {\color{black}{ L^6_{z}}} \Big) \bigg] +{ \color{blue} \pmb{a_s^6}} \bigg[  -  \bigg\{   \frac{16}{15} C_A^6 \bigg\} L_z^{11}  
       -\bigg\{ \frac{656}{135} C_A^5 n_f - \frac{5552}{135} C_A^6 \bigg\} L_z^{10}   +\bigg\{  - \frac{80}{9} C_A^4 n_f^2 + \frac{14252}{81} C_A^5 n_f - \bigg( \frac{58496}{81}  
       - \frac{272}{3} \zeta_2 \bigg) C_A^6 \bigg\}  L_z^9  +  \mathcal{O}\Big( {\color{black}{ L^8_{z}}} \Big) \bigg]
       + {\color{blue} \pmb{a_s^7}} \bigg[ -  \bigg\{ \frac{16}{45} C_A^7 \bigg\} L_z^{13} -  \bigg\{   \frac{296}{135} C_A^6 n_f 
       - \frac{2396}{135} C_A^7 \bigg\} L_z^{12} +  \bigg\{  - \frac{256}{45} C_A^5 n_f^2 + \frac{41864}{405} C_A^6 n_f - \bigg( \frac{164912}{405}  - \frac{128}{3} \zeta_2 \bigg) C_A^7 \bigg\}  L_z^{11} 
       +  \mathcal{O}\Big( {\color{black}{ L^{10}_{z}}} \Big) \Bigg]\,.
\end{autobreak}
\end{align}

%% file: appendix.tex
\section{NSV Partonic coefficient functions $\Delta^{NSV}_{c}$} \label{ap:del}
The partonic coefficient function given in Eq.(\ref{Delexp}) can be written as,
\begin{equation}
    \Delta_{c}^{(i)}(Q^2,\mu_R^2,\mu_F^2,z) =  \Delta_{c}^{SV,(i)}(Q^2,\mu_R^2,\mu_F^2,z) +  \Delta_{c}^{NSV,(i)}(Q^2,\mu_R^2,\mu_F^2,z)
\end{equation}
where $\Delta_{c}^{SV,(i)}(Q^2,\mu_R^2,\mu_F^2,z)$ can be found in \cite{Ravindran:2005vv,Ravindran:2006cg,deFlorian:2012za,Ahmed:2014cla,Li:2014afw}. The $\Delta_{c}^{NSV,(i)}$ to fourth order for DIS and SIA after setting $\mu_R^2 = \mu_F^2 = Q^2$ has the following expansion :
\begin{equation}
  \Delta_{c}^{NSV,(i)}(z) = \sum_{k=0}^{(2i-1)}\Delta_{c}^{ik}  \log^k (1-z)
\end{equation}

%The following results for DIS as well as SIA are provided in the ancillary files supplied with the \arXiv \  submission. We also put $\Delta^{30}_c,\cdots,\Delta^{32}_c$ and $\Delta^{40}_c,\cdots,\Delta^{45}_c$ in the ancillary files as they were lengthy. 

\begin{flushleft}

Below, we present the $\Delta_{c}^{ik}$ for DIS($c = q$ for photon exchange and $c = g$ for Higgs exchange).
\end{flushleft}

\begin{small}
\begin{align}

\begin{autobreak}

   \Delta_{c}^{10} =
        \overline{\varphi}_{c,1}^{(0)} \,,
        
\end{autobreak}
\nonumber \\

\begin{autobreak}
   
   \Delta_{c}^{11} =
        \overline{\varphi}_{c,1}^{(1)} 

       + D^c_1 \,,
       
\end{autobreak}
\nonumber \\

\begin{autobreak}
   
   \Delta_{c}^{12} =
         C^c_1 \,,
         
\end{autobreak}
\nonumber \\

\begin{autobreak}
   
   \Delta_{c}^{20} =
         \overline{\varphi}_{c,2}^{(0)}

       + \overline{\varphi}_{c,1}^{(1)}   \zeta_2   \bigg(  f^c_1 + B^c_1 \bigg) 

       + 2   \overline{\varphi}_{c,1}^{(0)}   \bigg(  \overline{\cal G}_{SV,1}^{c} + g^{c,1}_1 \bigg) 

       + D^c_1   \zeta_2   \bigg(  f^c_1 + B^c_1 \bigg) 

       - 2   C^c_1   \zeta_3   \bigg(  f^c_1 + B^c_1 \bigg) 

       + A^c_1 \overline{\varphi}_{c,1}^{(1)}   \zeta_3

       + A^c_1 D^c_1   \zeta_3

       - \frac{1}{5} A^c_1 C^c_1    \zeta_2^2 

       - \frac{1}{2} (A^c_1)^2   \zeta_2 

       + \frac{1}{2}   (f^c_1)^2 
       
       + B^c_1   f^c_1 
       
       + \frac{1}{2}   (B^c_1)^2 \,,
\end{autobreak}
\nonumber \\

\begin{autobreak}
   
   \Delta_{c}^{21} =
         \overline{\varphi}_{c,2}^{(1)}

       + 2 \overline{\varphi}_{c,1}^{(1)}   \bigg(  \overline{\cal G}_{SV,1}^{c} + g^{c,1}_1 \bigg) 

       - \overline{\varphi}_{c,1}^{(0)}   \bigg(  \beta_0 + f^c_1 + B^c_1 \bigg) 

       + D^c_2 

       + 2   D^c_1   \bigg(  \overline{\cal G}_{SV,1}^{c} + g^{c,1}_1 \bigg) 

       + 2   C^c_1   \zeta_2   \bigg( f^c_1 + B^c_1\bigg) 

       - A^c_1   \bigg( f^c_1  + B^c_1 \bigg) 

       - A^c_1 \overline{\varphi}_{c,1}^{(1)}   \zeta_2 

       - A^c_1 D^c_1   \zeta_2

       + 4   A^c_1   C^c_1   \zeta_3 \,,
       
\end{autobreak}
\nonumber \\

\begin{autobreak}
   
   \Delta_{c}^{22} =
         \overline{\varphi}_{c,2}^{(2)}

       - \overline{\varphi}_{c,1}^{(1)}   \bigg( \beta_0 + f^c_1 + B^c_1 \bigg) 

       - \frac{1}{2}   D^c_1   \bigg(  \beta_0 + 2   f^c_1 + 2   B^c_1 \bigg) 

       + C^c_2 

       + 2   C^c_1   \bigg( \overline{\cal G}_{SV,1}^{c} + g^{c,1}_1\bigg) 

       + \frac{1}{2} A^c_1 \overline{\varphi}_{c,1}^{(0)} 

       - 2   A^c_1   C^c_1   \zeta_2

       + \frac{1}{2} (A^c_1)^2  \,,
\end{autobreak}
\nonumber \\

\begin{autobreak}
   
   \Delta_{c}^{23} =
       -  \frac{1}{2}   C^c_1   \bigg(  \beta_0 + 2   f^c_1 + 2   B^c_1 \bigg) 

       + \frac{1}{2} A^c_1 \overline{\varphi}_{c,1}^{(1)} 

       + \frac{1}{2} A^c_1 D^c_1  \,,
\end{autobreak}
\nonumber \\

\begin{autobreak}
   
   \Delta_{c}^{24} =
        \frac{1}{2} A^c_1 C^c_1  \,,
\end{autobreak}
\nonumber \\

\begin{autobreak}
   
   \Delta_{c}^{33} =
         \overline{\varphi}_{c,3}^{(3)}

       - \overline{\varphi}_{c,2}^{(2)}   \bigg(  2 \beta_0 + f^c_1 + B^c_1 \bigg) 

       + \frac{1}{2} \overline{\varphi}_{c,1}^{(1)}   \bigg(  2   \beta_0^2 + 3   f^c_1 \beta_0 + (f^c_1)^2 + 3   B^c_1 \beta_0 + 2   B^c_1   f^c_1 + (B^c_1)^2 \bigg) 

       + \frac{1}{6} D^c_1   \bigg(  2 \beta_0^2 + 6 f^c_1 \beta_0 + 3 (f^c_1)^2 + 6 B^c_1 \beta_0 + 6 B^c_1 f^c_1 + 3 (B^c_1)^2 \bigg) 

       - C^c_2   \bigg( \beta_0 + f^c_1 + B^c_1 \bigg) 

       - \frac{1}{2} C^c_1   \bigg(\beta_1 + 6 \overline{\cal G}_{SV,1}^{c} \beta_0 + 2  g^{c,1}_1 \beta_0 + 2 f^c_2 + 4 f^c_1 \overline{\cal G}_{SV,1}^{c} + 4 f^c_1 g^{c,1}_1
          + 2 B^c_2 + 4 B^c_1 \overline{\cal G}_{SV,1}^{c} + 4 B^c_1 g^{c,1}_1 \bigg) 

       + \frac{1}{2} A^c_2 \overline{\varphi}_{c,1}^{(1)} 

       + \frac{1}{2} A^c_2 D^c_1 

       + \frac{1}{2} A^c_1 \overline{\varphi}_{c,2}^{(1)}  

       + A^c_1 \overline{\varphi}_{c,1}^{(1)}   \bigg(  \overline{\cal G}_{SV,1}^{c} + g^{c,1}_1 \bigg) 

       - \frac{1}{6} A^c_1 \overline{\varphi}_{c,1}^{(0)}    \bigg(  4 \beta_0 + 3 f^c_1 + 3 B^c_1 \bigg) 

       + \frac{1}{2} A^c_1 D^c_2 

       + A^c_1 D^c_1   \bigg(  \overline{\cal G}_{SV,1}^{c} + g^{c,1}_1 \bigg) 

       + \frac{1}{2} A^c_1 C^c_1   \zeta_2   \bigg(  5 \beta_0 + 8 f^c_1 + 8 B^c_1 \bigg) 

       - \frac{1}{2} (A^c_1)^2   \bigg(   \beta_0 + 2 f^c_1 + 2 B^c_1 \bigg) 

       - (A^c_1)^2 \overline{\varphi}_{c,1}^{(1)}   \zeta_2 

       - (A^c_1)^2 D^c_1   \zeta_2

       + 5   (A^c_1)^2 C^c_1   \zeta_3 \,,
       
\end{autobreak}
\nonumber \\

\begin{autobreak}
   
   \Delta_{c}^{34} =
         \frac{1}{6} C^c_1   \bigg(  2 \beta_0^2 + 6 f^c_1 \beta_0 + 3 (f^c_1)^2 + 6 B^c_1 \beta_0 + 6 B^c_1 f^c_1 + 3 (B^c_1)^2 \bigg) 

       + \frac{1}{2} A^c_2 C^c_1 

       + \frac{1}{2} A^c_1 \overline{\varphi}_{c,2}^{(2)} 

       - \frac{1}{6} A^c_1 \overline{\varphi}_{c,1}^{(1)}    \bigg(  4 \beta_0 + 3 f^c_1 + 3 B^c_1 \bigg) 

       - \frac{1}{12} A^c_1 D^c_1    \bigg(   5 \beta_0 + 6 f^c_1 + 6 B^c_1 \bigg) 

       + \frac{1}{2} A^c_1 C^c_2 

       + A^c_1 C^c_1   \bigg( \overline{\cal G}_{SV,1}^{c} + g^{c,1}_1 \bigg) 

       + \frac{1}{8} (A^c_1)^2 \overline{\varphi}_{c,1}^{(0)} 

       - \frac{3}{2} (A^c_1)^2 C^c_1  \zeta_2

       + \frac{1}{4} (A^c_1)^3  \,,
\end{autobreak}
\nonumber \\

\begin{autobreak}
   
   \Delta_{c}^{35} =
       - \frac{1}{12} A^c_1 C^c_1   \bigg(  5 \beta_0 + 6 f^c_1 + 6 B^c_1 \bigg) 

       + \frac{1}{8} (A^c_1)^2 \overline{\varphi}_{c,1}^{(1)} 

       + \frac{1}{8} (A^c_1)^2 D^c_1  \,,
\end{autobreak}
\nonumber \\

\begin{autobreak}
   
   \Delta_{c}^{36} =
        \frac{1}{8} (A^c_1)^2 C^c_1  \,,
\end{autobreak}
\nonumber \\

\begin{autobreak}
   
  \Delta_{c}^{45} =
       - \frac{1}{12} C^c_1    \bigg( 3 \beta_0^3 + 11 f^c_1 \beta_0^2 + 9 (f^c_1)^2 \beta_0 + 2 (f^c_1)^3 + 11 B^c_1 \beta_0^2 + 18 B^c_1 f^c_1 \beta_0 + 6 B^c_1 (f^c_1)^2 
       + 9 (B^c_1)^2 \beta_0 + 6 (B^c_1)^2 f^c_1 + 2 (B^c_1)^3 \bigg) 

       - \frac{1}{12} A^c_2 C^c_1  \bigg( 7 \beta_0 + 6 f^c_1 + 6 B^c_1 \bigg) 

       + \frac{1}{2} A^c_1 \overline{\varphi}_{c,3}^{(3)} 

       - \frac{1}{6} A^c_1 \overline{\varphi}_{c,2}^{(2)}   \bigg(  7 \beta_0 + 3 f^c_1 + 3 B^c_1 \bigg) 

       + \frac{1}{12} A^c_1 \overline{\varphi}_{c,1}^{(1)}   \bigg(  9 \beta_0^2 + 11 f^c_1 \beta_0 + 3 (f^c_1)^2 + 11 B^c_1 \beta_0 + 6 B^c_1 f^c_1 + 3 (B^c_1)^2 \bigg) 

       + \frac{1}{12}  A^c_1 D^c_1   \bigg( 4 \beta_0^2 + 8 f^c_1 \beta_0 + 3 (f^c_1)^2 + 8 B^c_1 \beta_0 + 6 B^c_1 f^c_1 + 3 (B^c_1)^2 \bigg) 

       - \frac{1}{6} A^c_1 C^c_2   \bigg( 4 \beta_0 + 3 f^c_1 + 3 B^c_1 \bigg) 

       - \frac{1}{12} A^c_1 C^c_1  \bigg( 5 \beta_1 + 22 \overline{\cal G}_{SV,1}^{c} \beta_0 + 10 g^{c,1}_1 \beta_0 + 6 f^c_2 + 12   f^c_1 \overline{\cal G}_{SV,1}^{c} + 12   f^c_1 g^{c,1}_1 + 6 B^c_2 + 12   B^c_1 \overline{\cal G}_{SV,1}^{c}
       
       + 12   B^c_1 g^{c,1}_1 \bigg) + \frac{1}{4} A^c_1 A^c_2 \overline{\varphi}_{c,1}^{(1)}

       + \frac{1}{4} A^c_1 A^c_2 D^c_1

       + \frac{1}{8} (A^c_1)^2 \overline{\varphi}_{c,2}^{(1)} 

       + \frac{1}{4} (A^c_1)^2 \overline{\varphi}_{c,1}^{(1)}   \bigg( \overline{\cal G}_{SV,1}^{c} + g^{c,1}_1 \bigg) 

       - \frac{1}{24} (A^c_1)^2 \overline{\varphi}_{c,1}^{(0)}   \bigg( 5 \beta_0 + 3 f^c_1 + 3 B^c_1 \bigg) 

       + \frac{1}{8} (A^c_1)^2 D^c_2 

       + \frac{1}{4} (A^c_1)^2 D^c_1   \bigg(\overline{\cal G}_{SV,1}^{c} + g^{c,1}_1 \bigg) 

       + \frac{1}{12} (A^c_1)^2 C^c_1   \zeta_2   \bigg(28 \beta_0 + 27 f^c_1 + 27 B^c_1 \bigg) 

       - \frac{1}{24} (A^c_1)^3   \bigg( 8 \beta_0 + 9 f^c_1 + 9 B^c_1 \bigg) 

       - \frac{3}{8} (A^c_1)^3 \overline{\varphi}_{c,1}^{(1)}  \zeta_2 

       - \frac{3}{8} (A^c_1)^3 D^c_1  \zeta_2 

       + \frac{7}{3} (A^c_1)^3 C^c_1  \zeta_3  \,,
       
\end{autobreak}
\nonumber \\

\begin{autobreak}
  
   \Delta_{c}^{46} =
        \frac{1}{12} A^c_1 C^c_1  \bigg(  4 \beta_0^2 + 8 f^c_1 \beta_0 + 3 (f^c_1)^2 
        + 8 B^c_1 \beta_0 + 6 B^c_1 f^c_1+ 3 (B^c_1)^2 \bigg) 

       + \frac{1}{4}  A^c_1 A^c_2 C^c_1  

       + \frac{1}{8} (A^c_1)^2 \overline{\varphi}_{c,2}^{(2)}  

       - \frac{1}{24} (A^c_1)^2 \overline{\varphi}_{c,1}^{(1)}   \bigg( 5 \beta_0 + 3 f^c_1 + 3 B^c_1 \bigg) 

       - \frac{1}{48} (A^c_1)^2 D^c_1   \bigg(  7 \beta_0 + 6 f^c_1 + 6 B^c_1 \bigg) 

       + \frac{1}{8} (A^c_1)^2 C^c_2 

       + \frac{1}{4} (A^c_1)^2 C^c_1   \bigg(  \overline{\cal G}_{SV,1}^{c} + g^{c,1}_1 \bigg) 

       + \frac{1}{48} (A^c_1)^3 \overline{\varphi}_{c,1}^{(0)} 

       -\frac{1}{2} (A^c_1)^3 C^c_1   \zeta_2 

       + \frac{1}{16} (A^c_1)^4 \,,
\end{autobreak}
\nonumber \\

\begin{autobreak}
  
   \Delta_{c}^{47} =
       - \frac{1}{48} (A^c_1)^2 C^c_1   \bigg( 7 \beta_0 + 6 f^c_1 + 6 B^c_1 \bigg) 

       + \frac{1}{48} (A^c_1)^3 \overline{\varphi}_{c,1}^{(1)}

       + \frac{1}{48} (A^c_1)^3 D^c_1  \,,
\end{autobreak}
\nonumber \\

\begin{autobreak}
     \Delta_{c}^{48} =
        \frac{1}{48} (A^c_1)^3 C^c_1  \,.
\end{autobreak}

\end{align}
\end{small}

\begin{flushleft}

We have put $\Delta^{30}_c,\cdots,\Delta^{32}_c$ and $\Delta^{40}_c,\cdots,\Delta^{44}_c$ along with the above results in the ancillary files as they are lengthy.
	Next, we present the $\Tilde{\Delta}_{c}^{ik}$ for SIA in terms of $\Delta_{c}^{ik}$ for DIS. The expansion coefficients $\overline{\varphi}_{c,i}^{(k)}$ should be replaced with $\Tilde{\varphi}_{c,i}^{(k)}$ given in Eq.(\ref{phieECt}) and Eq.(\ref{phieECL}), in the relations provided below.
\end{flushleft}

\begin{small}
\begin{flalign}

\begin{autobreak}

     \Tilde{\Delta}_{c}^{20} = 
         \Delta_{c}^{20} + 3 A^c_1 \zeta_2 \overline{\varphi}_{c,1}^{(0)} \,,

\end{autobreak}
%\nonumber
\nonumber \\

\begin{autobreak}

     \Tilde{\Delta}_{c}^{21} = 
         \Delta_{c}^{21} + 3 A^c_1 \zeta_2 \bigg( D^c_1 + \overline{\varphi}_{c,1}^{(1)}\bigg) \,,

\end{autobreak}
\nonumber \\

\begin{autobreak}

     \Tilde{\Delta}_{c}^{22} = 
         \Delta_{c}^{22} + 3 A^c_1 \zeta_2 C^c_1 \,,

\end{autobreak}
\nonumber \\

\begin{autobreak}

     \Tilde{\Delta}_{c}^{33} = 
         \Delta_{c}^{33} + \frac{3}{2} A^c_1 \zeta_2   \bigg( -C^c_1 \beta_0 - 2 f^c_1 C^c_1 - 2 B^c_1 C^c_1 + A^c_1 D^c_1 + A^c_1 \overline{\varphi}_{c,1}^{(1)} \bigg) \,,

\end{autobreak}
%\nonumber 
\nonumber \\

\begin{autobreak}

     \Tilde{\Delta}_{c}^{34} = 
         \Delta_{c}^{34} + \frac{3}{2} (A^c_1)^2 \zeta_2 C^c_1 \,,

\end{autobreak}
\nonumber \\

\begin{autobreak}

     \Tilde{\Delta}_{c}^{45} = 
         \Delta_{c}^{45} + \frac{1}{8} (A^c_1)^2 \zeta_2   \bigg( -10 C^c_1 \beta_0 - 12 f^c_1 C^c_1 - 12 B^c_1 C^c_1 + 3 A^c_1 D^c_1 + 3 A^c_1 \overline{\varphi}_{c,1}^{(1)} \bigg) \,,

\end{autobreak}
\nonumber \\
   
\begin{autobreak}

     \Tilde{\Delta}_{c}^{46} = 
         \Delta_{c}^{46} + \frac{3}{8} (A^c_1)^3 \zeta_2 C^c_1   \,.

\end{autobreak}

\end{flalign}
\end{small} 

\begin{flushleft}

The expressions for $\Tilde{\Delta}_{c}^{10}$, $\Tilde{\Delta}_{c}^{11}$, $\Tilde{\Delta}_{c}^{12}$, $\Tilde{\Delta}_{c}^{23}$, $\Tilde{\Delta}_{c}^{24}$, $\Tilde{\Delta}_{c}^{35}$, $\Tilde{\Delta}_{c}^{36}$, $\Tilde{\Delta}_{c}^{47}$ and $\Tilde{\Delta}_{c}^{48}$ are exactly same as the corresponding expressions for the case of DIS. The above results with the explicit dependence on $\mu_R^2$ and $\mu_F^2$ are provided in the ancillary files supplied with the arXiv submission. 
\end{flushleft}

\section{ SV Coefficients $\overline{\cal G}_{SV,i}^{c}$ in $\Delta_c$} \label{ap:gsv}

We present here the expressions for $\overline{\cal G}_{SV,i}^{c}$ used in the $\Delta_{c}^{ik}$ for DIS.

\begin{small}
\begin{align}

\begin{autobreak}

\overline{\cal G}_{SV,1}^{c} = 
       \overline{\cal G}_1^{c,(1)} \,,
\end{autobreak}
\nonumber \\

\begin{autobreak}
\overline{\cal G}_{SV,2}^c =
       \bigg( \overline{\cal G}_2^{c,(1)} + 2 \beta_0 \overline{\cal G}_1^{c,(2)} \bigg) \,,
\end{autobreak}
\nonumber \\

\begin{autobreak}
\overline{\cal G}_{SV,3}^c =
       \bigg( \frac{2}{3} \overline{\cal G}_3^{c,(1)} + \frac{4}{3} \beta_1 \overline{\cal G}_1^{c,(2)} + \frac{4}{3} \beta_0 \overline{\cal G}_2^{c,(2)} + \frac{2}{3} \beta_0^2 \overline{\cal G}_1^{c,(3)} \bigg) \,,
\end{autobreak}
\nonumber \\

\begin{autobreak}

\overline{\cal G}_{SV,4}^c =
       \bigg( \frac{1}{2} \overline{\cal G}_4^{c,(1)} + \beta_2 \overline{\cal G}_1^{c,(2)} + \beta_1 \overline{\cal G}_2^{c,(2)} + \beta_0 \overline{\cal G}_3^{c,(2)} + 2 \beta_0^2 \overline{\cal G}_2^{c,(3)} + 4 \beta_0^3 \overline{\cal G}_1^{c,(4)} + 4 \beta_0 \beta_1 \overline{\cal G}_1^{c,(3)} \bigg) \,,

\end{autobreak}
\end{align}
\end{small}

\begin{flushleft}
The explicit results for $\overline{\cal G}_{SV,1}^{c}$, $\overline{\cal G}_{SV,2}^{c}$ and $\overline{\cal G}_{SV,3}^{c}$ in terms of colour factors $C_A$, $C_F$ and $n_f$ are given in the ancillary files provided with the \arXiv \ submission.
\end{flushleft}

\section{NSV resummation constants  $h^c_{ij}(\omega)$} \label{ap:hij}

The resummation constants $h^c_{ij}(\omega)$ given in Eq.(\ref{hg}) are found to be as following. Here $\bar{L}_{\omega}=\log(1-\omega)$, $L_{qr} = \log(\frac{Q^2}{\mu_R^2})$, $L_{fr} = \log(\frac{\mu_F^2}{\mu_R^2})$ and $\omega = \beta_0 as(\mu_R^2) \log N$.

\begin{small}
\begin{align}

\begin{autobreak}

h^c_{00}(\omega) =
         \frac{1}{\beta_0} \overline{L}_{\omega} \bigg[
           \gamma^B_1 D^c_1
          - \gamma^B_2 C^c_1
          \bigg] \,,
\end{autobreak}
\nonumber \\

\begin{autobreak}

h^c_{01}(\omega) =
           \frac{1}{\beta_0} \overline{L}_{\omega}  \bigg[
          - \gamma^B_1 C^c_1
          \bigg] \,,
\end{autobreak}
\nonumber \\

\begin{autobreak}
 h^c_{10}(\omega) =
         \frac{1}{1-\omega} \Bigg[ 
             \overline{\varphi}_{c,1}^{(0)} \gamma^B_1
           - \overline{\varphi}_{c,1}^{(1)} \gamma^B_2
           - \frac{D^c_2}{\beta_0} 
             \gamma^B_1 \omega
           + \frac{\beta_1 D^c_1}{\beta_0^2} \gamma^B_1 \bigg \{
                 \omega
               + \overline{L}_{\omega} \bigg \}
           - D^c_1  \bigg \{
                 \gamma^B_2
               - \gamma^B_1 L_{qr}
               + \gamma^B_1 L_{fr}
               - \gamma^B_1 L_{fr} \omega
                 \bigg \}
           + \frac{C^c_2}{\beta_0}  \bigg \{
                 \gamma^B_2 \omega
                 \bigg \}

           - \frac{\beta_1 C^c_1}{\beta_0^2} \gamma^B_2 \bigg \{
                  \omega
               +  \overline{L}_{\omega}
                 \bigg \}

           + C^c_1  \bigg \{ 
                2 \gamma^B_3
               - \gamma^B_2 L_{qr}
               + \gamma^B_2 L_{fr}
               - \gamma^B_2 L_{fr} \omega
               \bigg \}
          \Bigg] \,,

\end{autobreak}
\nonumber \\

\begin{autobreak}
 
h^c_{11}(\omega) = 
        \frac{1}{1-\omega} \Bigg[ 

            - \overline{\varphi}_{c,1}^{(1)} 
             \gamma^B_1

            + \frac{C^c_2}{\beta_0}  \bigg \{ 
               \gamma^B_1 \omega
              \bigg \}

            - \frac{\beta_1 C^c_1}{\beta_0^2} \gamma^B_1 \bigg \{
                 \omega
              +  \overline{L}_{\omega}
              \bigg \}

            + C^c_1  \bigg \{ 
                \gamma^B_2
              - \gamma^B_1 L_{qr}
              + \gamma^B_1 L_{fr}
              - \gamma^B_1 L_{fr} \omega
              \bigg \}
          
        \Bigg]  \,,
\end{autobreak}
\nonumber \\

\begin{autobreak}

 h^c_{21}(\omega) =
        \frac{1}{(1-\omega)^2} \Bigg[

          \frac{\beta_1}{\beta_0} \overline{\varphi}_{c,1}^{(1)}  \bigg \{
                \gamma^B_1 \overline{L}_{\omega}
          \bigg \}

          +  2  \overline{\varphi}_{c,2}^{(2)}  \gamma^B_2

          -     \overline{\varphi}_{c,2}^{(1)}  \gamma^B_1

          -     \beta_0 \overline{\varphi}_{c,1}^{(1)}  \bigg \{
         
                     \gamma^B_2
         
                  -  \gamma^B_1 L_{qr}
                \bigg \}

          +    \frac{C^c_3}{2 \beta_0} \gamma^B_1  \bigg \{
                       2 \omega
                    -    \omega^2
                       \bigg \}

          -     \frac{\beta_1 C^c_2}{2 \beta_0^2} \gamma^B_1 \bigg \{
                      2 \omega
                      - \omega^2
                      + 2 \overline{L}_{\omega}
                      \bigg \}

          +     C^c_2   \bigg \{
                      \gamma^B_2
                    - \gamma^B_1 L_{qr}
                    + \gamma^B_1 L_{fr}
                    - 2 \gamma^B_1 L_{fr} \omega
                    + \gamma^B_1 L_{fr} \omega^2
                        \bigg \}

          -     \frac{\beta_1^2 C^c_1}{2 \beta_0^3} \gamma^B_1 \bigg \{
                      \omega^2
                - \overline{L}_{\omega}^2
                      \bigg \}

          +     \frac{\beta_2 C^c_1}{2 \beta_0^2}   \bigg \{
                       \gamma^B_1 \omega^2
                      \bigg \}
          
          -     \frac{\beta_1 C^c_1}{\beta_0}  \bigg \{
                      \gamma^B_2 \overline{L}_{\omega}
                    - \gamma^B_1 L_{qr} \overline{L}_{\omega}
                      \bigg \}

          +    \frac{C^c_1 \beta_0}{2}   \bigg \{
                    2  \gamma^B_3
                    -2 \gamma^B_2 L_{qr}
                    +  \gamma^B_1 L_{qr}^2
                    -  \gamma^B_1 L_{fr}^2
                    + 2 \gamma^B_1 L_{fr}^2 \omega
                    - \gamma^B_1 L_{fr}^2 \omega^2
                    \bigg \}
          
          \Bigg] \,,

\end{autobreak}
\nonumber \\

\begin{autobreak}

h^c_{22}(\omega) =
       \frac{1}{(1-\omega)^2} \Bigg[
            \overline{\varphi}_{c,2}^{(2)} \gamma^B_1
        \Bigg] \,,

\end{autobreak}
\nonumber \\

\begin{autobreak}

 h^c_{32}(\omega) =
         \frac{1}{(1-\omega)^3} \Bigg[
            \frac{\beta_1}{\beta_0} \overline{\varphi}_{c,2}^{(2)}  \bigg \{
                 - 2 \gamma^B_1 \overline{L}_{\omega}
                 \bigg \}

                 - 3 \overline{\varphi}_{c,3}^{(3)}
                     \gamma^B_2

                 +  \overline{\varphi}_{c,3}^{(2)}
                    \gamma^B_1

                 + 2 \beta_0 \overline{\varphi}_{c,2}^{(2)}  \bigg \{
                        \gamma^B_2
                     -  \gamma^B_1 L_{qr}
                    \bigg \}
            \Bigg] \,,        

\end{autobreak}
\nonumber \\

\begin{autobreak}

h^c_{33}(\omega) = 
         \frac{1}{(1-\omega)^3} \Bigg[
           - \overline{\varphi}_{c,3}^{(3)} 
             \gamma^B_1
         \Bigg] \,,
         
\end{autobreak}
\nonumber \\

\begin{autobreak}

     h^c_{42}(\omega) =
       \frac{1}{(1-\omega)^4} \Bigg[
       
         \frac{\beta_1^2}{\beta_0^2} \overline{\varphi}_{c,2}^{(2)}   \bigg \{
          - 2 \gamma^B_1 \omega
          - 2 \gamma^B_1 \overline{L}_{\omega}
          + 3 \gamma^B_1 \overline{L}_{\omega}^2
          \bigg \}

       + \frac{2 \beta_2}{\beta_0} \overline{\varphi}_{c,2}^{(2)}
          \gamma^B_1 \omega

       +  \frac{9 \beta_1}{\beta_0} \overline{\varphi}_{c,3}^{(3)}
          \gamma^B_2 \overline{L}_{\omega}

       -  \frac{3 \beta_1}{\beta_0} \overline{\varphi}_{c,3}^{(2)}
            \gamma^B_1 \overline{L}_{\omega}

       + 12 \overline{\varphi}_{c,4}^{(4)} 
          \gamma^B_3

       - 3 \overline{\varphi}_{c,4}^{(3)} 
           \gamma^B_2

       +  \overline{\varphi}_{c,4}^{(2)}  
          \gamma^B_1

       + 2 \beta_1 \overline{\varphi}_{c,2}^{(2)}  \bigg \{
             \gamma^B_2
          - 3 \gamma^B_2 \overline{L}_{\omega}
          -  \gamma^B_1 L_{qr}
          + 3 \gamma^B_1 L_{qr} \overline{L}_{\omega}
          \bigg \}

       - 9 \beta_0 \overline{\varphi}_{c,3}^{(3)}  \bigg \{
           2 \gamma^B_3
          -  \gamma^B_2 L_{qr}
          \bigg \}  
       
       + 3 \beta_0 \overline{\varphi}_{c,3}^{(2)}  \bigg \{
              \gamma^B_2
          -   \gamma^B_1 L_{qr}
          \bigg \}

       + 3 \beta_0^2 \overline{\varphi}_{c,2}^{(2)} \bigg \{
            2 \gamma^B_3
          - 2 \gamma^B_2 L_{qr}
          +  \gamma^B_1 L_{qr}^2
          \bigg \}
          
        \Bigg] \,,

\end{autobreak}
\nonumber \\

\begin{autobreak}

 h^c_{43}(\omega) =
        \frac{1}{(1-\omega)^4} \bigg[

         \frac{3 \beta_1}{\beta_0} \overline{\varphi}_{c,3}^{(3)} 
          \gamma^B_1 \overline{L}_{\omega}

       + 4 \overline{\varphi}_{c,4}^{(4)}   
           \gamma^B_2

       - \overline{\varphi}_{c,4}^{(3)}
          \gamma^B_1

       - 3 \beta_0 \overline{\varphi}_{c,3}^{(3)}  \bigg \{
           \gamma^B_2
          - \gamma^B_1 L_{qr}
          \bigg \}
          
          \bigg] \,,

\end{autobreak}
\nonumber \\

\begin{autobreak}

  h^c_{44}(\omega) =

       \frac{1}{(1-\omega)^4} \Bigg[
       \overline{\varphi}_{c,4}^{(4)} \gamma^B_1
       \Bigg] \,.
          
\end{autobreak}

\end{align}   
\end{small}

\begin{flushleft}

The above results along with the bigger ones $(h^c_{20}(\omega), h^c_{30}(\omega),h^c_{31}(\omega))$ and 
$( h^c_{40}(\omega),h^c_{41}(\omega))$ are all provided in the ancillary files with the \arXiv \  submission.

\end{flushleft}  

\section{NSV resummation constants $\bar g^c_{i}(\omega)$ } \label{ap:gbar} 

The resummation constants $\bar g^c_{i}(\omega)$ given in Eq.(\ref{hg}) are presented below. Here $\bar{L}_{\omega}=\log(1-\omega)$, $L_{qr} = \log(\frac{Q^2}{\mu_R^2})$, $L_{fr} = \log(\frac{\mu_F^2}{\mu_R^2})$ and $\omega = \beta_0 a_s(\mu_R^2) \log(N)$. Also, $\textbf{B}_{DIS,i}^c$ are the threshold exponent given in \cite{Ravindran:2005vv}. 

\begin{small}
\begin{align}

\begin{autobreak}
   \overline{g}^c_2(\omega) =

          \frac{A^c_1}{2 \beta_0} 
          \overline{L}_{\omega} \,,
          
\end{autobreak}
\nonumber \\

\begin{autobreak}
   \overline{g}^c_3(\omega) =  
    \frac{1}{(1-\omega)} \bigg[
        - \frac{A^c_2 \omega}{2 \beta_0} 

       +  \frac{\beta_1 A^c_1}{2 \beta_0^2} \bigg \{
            \omega
          + \overline{L}_{\omega}
          \bigg \}

       - \frac{A^c_1}{2}   \bigg \{
             1
          +  \gamma_E
          -  L_{qr}
          +  L_{fr}
          -  L_{fr} \omega
          \bigg \}

       + \frac{\bold B_{DIS,1}^{c}}{2}
    \bigg] \,,

\end{autobreak}
\nonumber \\

\begin{autobreak}
   
 \overline{g}^c_4(\omega) =
        \frac{1}{(1-\omega)^2} \bigg[
        
        - \frac{\beta_1}{2 \beta_0}  
          \bold B_{DIS,1}^{c}  \overline{L}_{\omega}

       + \frac{\beta_0}{2} \bold B_{DIS,1}^{c} \bigg \{
            1 
          + \gamma_E
          -  L_{qr}
          \bigg \}

       - \frac{A^c_3}{4 \beta_0} \bigg \{          
           2 \omega
          - \omega^2
          \bigg \}

       +  \frac{\beta_1 A^c_2}{4 \beta_0^2}  \bigg \{
           2 \omega
          - \omega^2
          + 2 \overline{L}_{\omega}
          \bigg \}
          
       - \frac{A^c_2}{2}   \bigg \{
          1
         +  \gamma_E
          - L_{qr}
          + L_{fr}
          - 2 L_{fr} \omega
          + L_{fr} \omega^2
          \bigg \}

       +  \frac{\beta_1^2 A^c_1}{4 \beta_0^3}  \bigg \{
            \omega^2
          -  \overline{L}_{\omega}^2
          \bigg \}

       -  \frac{\beta_2 A^c_1}{4 \beta_0^2}  \bigg \{
           \omega^2
          \bigg \}
          
        + \frac{\beta_1  A^c_1}{2 \beta_0} \bigg \{
           \overline{L}_{\omega}
          + \overline{L}_{\omega} \gamma_E
          - L_{qr} \overline{L}_{\omega}
          \bigg \}

       -  \frac{\beta_0 A^c_1}{4}  \bigg \{
           2\gamma_E
          +  \gamma_E^2
          +  \zeta_2
          - 2 L_{qr}
          - 2 L_{qr} \gamma_E
          + L_{qr}^2
          - L_{fr}^2
          + 2 L_{fr}^2 \omega
          - L_{fr}^2 \omega^2
          \bigg \}

       + \frac{\bold B_{DIS,2}^{c}}{2}
        
        \bigg] \,,   
   
\end{autobreak}
\nonumber \\

\begin{autobreak}

 \overline{g}^c_5(\omega) =
        \frac{1}{(1-\omega)^3} \bigg[

       -  \frac{\beta_1^2 \bold B_{DIS,1}^{c}}{2 \beta_0^2} \bigg \{
          \omega 
          + \overline{L}_{\omega} 
          - \overline{L}_{\omega}^2 
          \bigg \}

       + \frac{\beta_2}{2 \beta_0} 
            \omega \bold B_{DIS,1}^{c}

       - \frac{\beta_1}{\beta_0} 
          \overline{L}_{\omega} \bold B_{DIS,2}^{c}

       + \frac{\beta_1 \bold B_{DIS,1}^{c}}{2}  \bigg \{
           1
          +  \gamma_E
          - 2 \overline{L}_{\omega} 
          - 2 \overline{L}_{\omega}  \gamma_E
          - L_{qr} 
          + 2 L_{qr} \overline{L}_{\omega} 
          \bigg \}
          
         + \beta_0 \bold B_{DIS,2}^{c}  \bigg \{
            1 
          + \gamma_E
          - L_{qr}
          \bigg \}

       + \frac{\beta_0^2 \bold B_{DIS,1}^{c}}{2} \bigg \{
            2 \gamma_E
          +   \gamma_E^2
          +   \zeta_2
          - 2 L_{qr} 
          - 2 L_{qr}  \gamma_E
          + L_{qr}^2 
          \bigg \}

       - \frac{A^c_4}{6 \beta_0}  \bigg \{
            3 \omega
          - 3 \omega^2
          + \omega^3
          \bigg \}
          
        + \frac{\beta_1 A^c_3}{6 \beta_0^2}  \bigg \{
            3 \omega
          - 3 \omega^2
          +  \omega^3
          + 3 \overline{L}_{\omega}
          \bigg \}

       - \frac{A^c_3}{2}  \bigg \{
            1
          + \gamma_E
          -  L_{qr}
          +  L_{fr}
          - 3 L_{fr} \omega
          + 3 L_{fr} \omega^2
          - L_{fr} \omega^3
          \bigg \}

       + \frac{\beta_1^2 A^c_2}{6 \beta_0^3}  \bigg \{
            3 \omega^2
          -  \omega^3
          - 3 \overline{L}_{\omega}^2
          \bigg \}
          
         - \frac{\beta_2 A^c_2}{6 \beta_0^2}  \bigg \{
            3 \omega^2
          -  \omega^3
          \bigg \}

       + \frac{\beta_1 A^c_2 \overline{L}_{\omega}}{\beta_0}  \bigg \{
            1
          + \gamma_E
          - L_{qr} 
          \bigg \}
        
       - \frac{A^c_2 \beta_0}{2}   \bigg \{
            2 \gamma_E
          +  \gamma_E^2
          +  \zeta_2
          - 2 L_{qr}
          - 2 L_{qr} \gamma_E
          + L_{qr}^2
          - L_{fr}^2
          + 3 L_{fr}^2 \omega
          - 3 L_{fr}^2 \omega^2
          + L_{fr}^2 \omega^3
          \bigg \}

       - \frac{\beta_1^3 A^c_1}{12 \beta_0^4}  \bigg \{
            3 \omega^2
          - 2 \omega^3
          + 6 \overline{L}_{\omega} \omega
          + 3 \overline{L}_{\omega}^2
          - 2 \overline{L}_{\omega}^3
          \bigg \}
          
        + \frac{\beta_1 \beta_2 A^c_1}{6 \beta_0^3}  \bigg \{
            3 \omega^2
          - 2 \omega^3
          + 3 \overline{L}_{\omega} \omega
          \bigg \}

       -  \frac{\beta_3  A^c_1}{12 \beta_0^2}  \bigg \{
             3 \omega^2
          -  2 \omega^3
          \bigg \}
          
        + \frac{\beta_1^2 A^c_1}{2 \beta_0^2}  \bigg \{
            \omega
          + \omega \gamma_E
          + \overline{L}_{\omega}
          + \overline{L}_{\omega} \gamma_E
          - \overline{L}_{\omega}^2
          - \overline{L}_{\omega}^2 \gamma_E
          - L_{qr} \omega
          - L_{qr} \overline{L}_{\omega}
          + L_{qr} \overline{L}_{\omega}^2
          \bigg \}

       -  \frac{\beta_2 A^c_1}{2 \beta_0}  \bigg \{
            \omega
          + \omega \gamma_E
          - L_{qr} \omega
          \bigg \}
        
        - \frac{A^c_1 \beta_1}{4}  \bigg \{
            2 \gamma_E
          + \gamma_E^2
          + \zeta_2
          - 4 \overline{L}_{\omega} \gamma_E
          - 2 \overline{L}_{\omega} \gamma_E^2
          - 2 \overline{L}_{\omega} \zeta_2
          - 2 L_{qr}
          - 2 L_{qr} \gamma_E
          + 4 L_{qr} \overline{L}_{\omega}
          + 4 L_{qr} \overline{L}_{\omega} \gamma_E
          + L_{qr}^2
          - 2 L_{qr}^2 \overline{L}_{\omega}
          - L_{fr}^2
          + 3 L_{fr}^2 \omega
          - 3 L_{fr}^2 \omega^2
          + L_{fr}^2 \omega^3
          \bigg \}
          
        - \frac{A^c_1 \beta_0^2}{6}  \bigg \{
            3 \gamma_E^2
          +  \gamma_E^3
          + 2 \zeta_3
          + 3 \zeta_2
          + 3 \zeta_2 \gamma_E
          - 6 L_{qr} \gamma_E
          - 3 L_{qr} \gamma_E^2
          - 3 L_{qr} \zeta_2
          + 3 L_{qr}^2
          + 3 L_{qr}^2 \gamma_E
          -  L_{qr}^3
          +  L_{fr}^3
          - 3 L_{fr}^3 \omega
          + 3 L_{fr}^3 \omega^2
          -  L_{fr}^3 \omega^3
          \bigg \}

       + \frac{\bold B_{DIS,3}^{c}}{2}
    
    \bigg] \,.

\end{autobreak}

\end{align}
\end{small}

\begin{flushleft}

As before here also we provide the above results along with $\bar g^c_{6}(\omega)$ in the ancillary files with the \arXiv \ submission.
 
\end{flushleft}